\documentclass[11pt]{article}

\usepackage[utf8]{inputenc}
\usepackage[T1]{fontenc}
\usepackage[margin=1in]{geometry}
\usepackage{amsmath, amssymb, amsfonts}
\usepackage{graphicx}
\graphicspath{{figures/}}
\usepackage{booktabs}
\usepackage{microtype}
\usepackage{xcolor}
\usepackage[round,authoryear]{natbib}
\usepackage[breaklinks=true,hidelinks]{hyperref}
\usepackage{tikz-cd}
\usepackage{tikz}
\usepackage{float}
\usepackage{multirow}
\usepackage{longtable}
\usepackage{array}
\usepackage{enumitem}
\usepackage{titlesec}
\usepackage[font=small,skip=4pt]{caption}
\usepackage{authblk}

\hypersetup{
  pdftitle={Beyond Object-Level Alignment: Do Brains and DNNs Preserve the Same Transformations?},
  pdfauthor={Yukiyasu Kamitani},
  pdfcreator={pdfTeX},
  pdfproducer={pdfTeX}
}
\newcommand{\nonviable}[1]{#1$^{\dagger}$}

\titlespacing*{\section}{0pt}{9pt plus 2pt minus 1pt}{4pt plus 1pt minus 1pt}
\titlespacing*{\subsection}{0pt}{7pt plus 2pt minus 1pt}{3pt plus 1pt minus 1pt}
\titlespacing*{\paragraph}{0pt}{4pt plus 1pt minus 1pt}{0.6em}

\title{Beyond Object-Level Alignment:\\Do Brains and DNNs Preserve the Same Transformations?}

\author[1,2]{Yukiyasu Kamitani}
\affil[1]{Graduate School of Informatics, Kyoto University}
\affil[2]{ATR Computational Neuroscience Laboratories}
\date{}

\begin{document}
\raggedbottom
\maketitle

\begin{abstract}
Brain--DNN alignment is usually assessed through stimulus-level correspondence or stimulus-set geometry. Inspired by category theory, we operationalize a different question: do brain and model preserve the same \emph{candidate transformations} among stimuli? We formalize this as approximate \emph{naturality} --- if a proxy-defined stimulus change is propagated through the brain side and then translated to the model side, the result should match translating first and then propagating through the model side, so that the naturality square approximately commutes. We quantify deviations from commutativity by a \emph{Naturality Violation Score} (NVS) normalized to a permutation null, shifting alignment from per-stimulus sameness to preservation of structure under an explicitly chosen comparison map. As a proof of concept, a controlled five-factor synthetic setting shows that NVS separates complementary alignment failures that aggregate object- and geometry-level scalars cannot resolve. Applied to fMRI responses from the GOD dataset (5 subjects), 3 vision DNNs, and 3 external embedding spaces used as limited World-Model proxies, the axis-resolved analysis reveals a hierarchy crossover: semantic axes align most strongly toward HVC and deeper DNN layers (animacy has the lowest pooled NVS among the tested axes, $\mathrm{NVS}^{\mathrm{animacy}} = 0.39$, compared with $0.52$ for the next-best axis and $1.0$ for the permutation-null baseline), whereas low- and mid-level visual axes align toward earlier visual cortex and shallower layers. Supporting analyses --- a 15-axis appendix atlas, dissociation tests against RSA/CKA and encoding/decoding accuracy, and a W-less anchor-ablation control --- support that the alignment is selective over candidate morphism families rather than uniform. NVS thereby turns brain--DNN comparison into a test of jointly preserved candidate transformations, relative to an explicitly chosen proxy space and permutation null, opening a path to richer proxy spaces and controlled world-side transformations.
\end{abstract}

\section{Introduction}
\label{sec:intro}

When brain and model assign similar codes to the same image, do they also \emph{move} between images in the same way? Existing alignment metrics --- encoding \citep{yamins2014} and decoding \citep{horikawa2017} accuracies, Brain-Score \citep{schrimpf2018}, the Brain Hierarchy (BH) score \citep{nonaka2021}, RSA \citep{kriegeskorte2008,kriegeskorte2015}, CKA \citep{kornblith2019}, and Procrustes alignment --- summarize how closely the two systems agree on individual stimuli or on overall stimulus-set geometry. In their standard use, these metrics do not directly test whether a specified stimulus transformation propagates compatibly through both systems: two systems can agree on per-stimulus predictions and on stimulus-set geometry yet still disagree about which changes among stimuli they treat as equivalent, stable, or meaningful.

\begin{figure}[!t]
  \centering
  \includegraphics[width=0.88\linewidth]{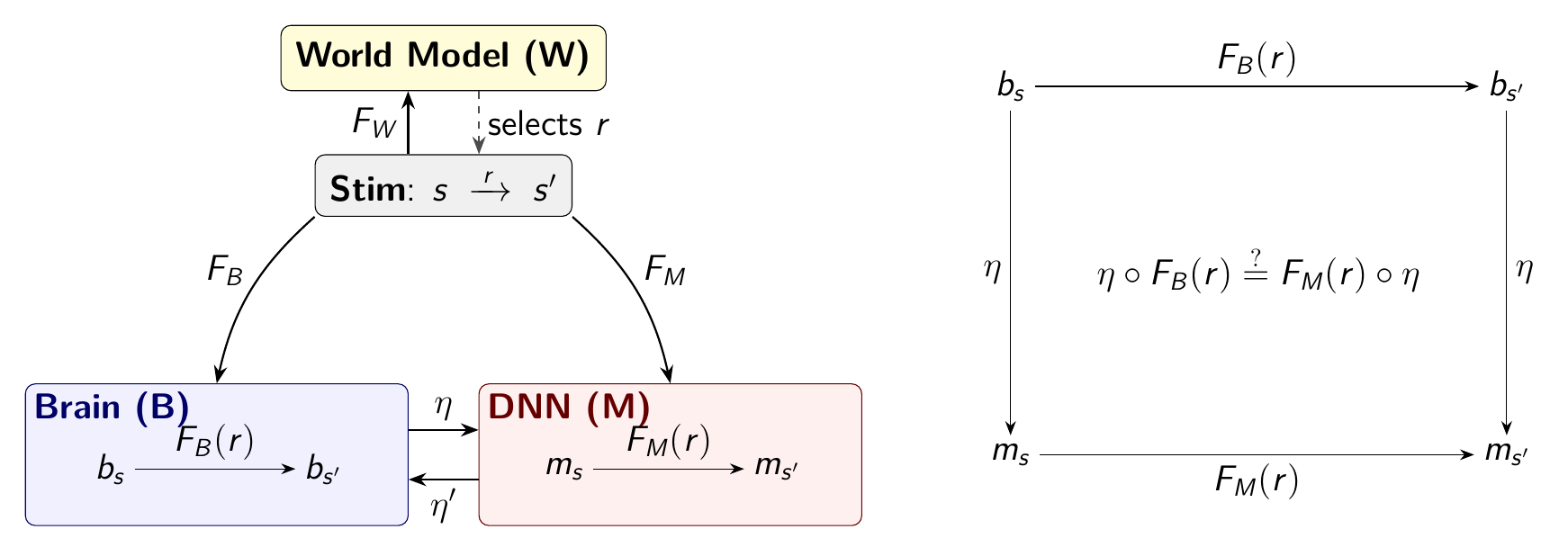}
  \caption{\textbf{Brain--DNN alignment as approximate naturality (motivational concept diagram).} \emph{Left:} the World Model $W$ parameterizes which stimulus change $r\colon s\to s'$ is tested; the translators $\eta\colon B\to M$ and $\eta'\colon M\to B$ connect the two sides. \emph{Right:} the naturality square (the analogous square is also asked of $\eta'$). This is the motivational categorical reading; in practice $F_B(r), F_M(r)$ are not estimated as endomorphisms of $B,M$ but realized at first order via the proxy chain $F_W, \Phi_B, \Phi_M$ (Fig.~\ref{fig:nvs-triangle}); $F_B, F_M$ are not enforced to be strict functors (App.~\ref{app:cat}).}
  \label{fig:cospan}
\end{figure}

Inspired by category theory, we frame this as a question about approximate \emph{naturality}: whether brain and model preserve the same proxy-defined candidate changes under translation, so that the two paths around the square approximately commute (Fig.~\ref{fig:cospan}; full categorical setup in App.~\ref{app:cat}). We use this as an \emph{operational analogy} rather than as a claim that the fitted maps constitute strict functors --- the tested operator families are empirical linear approximations defined on sampled stimulus pairs. This shifts alignment from objectwise sameness to preservation of structure under an explicitly chosen comparison map. Concretely, $B$ (brain) and $M$ (model) are ambient vector spaces with stimulus points $b_s\in B$, $m_s\in M$. A tested stimulus change $r\colon s\to s'$ is realized on each side by linear operators $F_B(r)\colon B\to B$ and $F_M(r)\colon M\to M$ (Fig.~\ref{fig:cospan}, left), which we approximate per edge from sampled stimulus pairs (formalized in \S\ref{sec:method}). We call the cross-system maps $\eta\colon B\to M$ and $\eta'\colon M\to B$ \emph{translators} \citep{shirakawa2025}: linear maps between brain activity and a target latent representation, with $\eta$ a decoder and $\eta'$ an encoder. In the present framework they act as approximate intertwiners.
For readability we write the forward square explicitly,
\begin{equation}
\eta\circ F_B(r) \;=\; F_M(r)\circ \eta,
\label{eq:naturality}
\end{equation}
i.e., the two paths --- propagate then translate, vs.\ translate then propagate --- should agree (Fig.~\ref{fig:cospan}, right). The same square is asked of $\eta'\colon M\to B$ in the reverse direction ($\eta'\circ F_M(r) = F_B(r)\circ\eta'$). Standard alignment metrics test only $\eta(b_s)\approx m_s$ at each stimulus individually; approximate naturality additionally constrains the \emph{change structure} --- whether the same candidate transformation is preserved across the two systems.

The candidate morphisms $r$ are specified through an external \emph{comparison space} $W$: the proxy through which the same candidate stimulus change is mapped into both brain and model spaces, sitting at the apex of the cospan that links them (Fig.~\ref{fig:cospan}; full categorical setup in App.~\ref{app:cat}). The choice of $W$ determines which transformations can be tested and is therefore part of the scientific question, not a nuisance parameter. In this paper, $W$ is a \emph{limited} World Model proxy in the broad neuro-AI sense \citep{ha2018,bisk2020}, given by an embedding function $F_W\colon\mathbf{Stim}\to W$; we instantiate three: CLIP-text \citep{radford2021}, DINOv2 \citep{oquab2023}, and DreamSim \citep{fu2023}, capturing language-grounded semantics, self-supervised visual structure, and human perceptual similarity, respectively. We do not treat these as full generative or causal world models, only as embedding spaces that supply a structured family of candidate morphisms. Differences within $W$ parametrize world-structural changes via the \emph{linear representation hypothesis} \citep{mikolov2013,park2024} --- the empirical observation that semantically meaningful features are encoded as approximately linear directions in modern embedding spaces, so that a single direction in $W$ corresponds to one such change. Restricting to a single \emph{concept axis} via a CAV \citep{kim2018} yields an axis-resolved version (formalized in \S\ref{sec:method}; throughout this paper, ``axis'' refers only to this CAV-based parameterization). The Naturality Violation Score (NVS) is defined in \S\ref{sec:method} as the relative $L^2$ residual of Eq.~\eqref{eq:naturality}, evaluated in both translation directions ($\eta, \eta'$) and normalized to a permutation null.

We evaluate this in two stages. A five-factor synthetic toy world serves as a \emph{proof of concept} (\S\ref{sec:poc}): morphism-level NVS recovers complementary failures that aggregate object- and geometry-level scalars collapse. On empirical data --- GOD fMRI from 5 subjects, 3 vision DNNs, and 3 World Model proxies (\S\ref{sec:results}) --- we first fit conventional per-stimulus encoding/decoding mappings and only then evaluate naturality post hoc, without optimizing for commutativity. The axis-resolved analysis reveals a \emph{hierarchy crossover} (Tab.~\ref{tab:hierarchy-crossover}): low-level photometric morphisms align at V1$\times$shallow cells, semantic morphisms at HVC$\times$deep cells, and mid-level morphisms in between, with animacy the strongest of the six main axes.

To keep the main narrative readable while preserving evidential support, the appendix is organized by scientific function rather than chronology: formal setup and data-processing details (Apps.~\ref{app:cat}--\ref{app:methods}), cross-subject reproducibility and confirmatory tests of the main claim (App.~\ref{app:primary-results}), atlas-wide $\mathrm{NVS}^a$ on the full 15-axis set (App.~\ref{app:atlas15}), the W-less control (App.~\ref{app:wless-section}), readout diagnostics and the variance decomposition (App.~\ref{app:model-diag}), alternative-explanation checks (App.~\ref{app:robust}), and metric-reduction / reproducibility material (Apps.~\ref{app:reductions}--\ref{app:repro}).

The scope is deliberately limited: empirical results come from $n=5$ subjects on a single dataset, depend on the chosen proxy space $W$ (and the embedding $F_W$ that defines it), and use category-theoretic naturality as an operational analogy rather than a strict functorial claim --- the operational maps used to realize each candidate morphism are fitted independently per edge without enforced composition (\S\ref{sec:method}; App.~\ref{app:cat}).

\textbf{Contributions:} (i) we reframe brain--DNN alignment as approximate naturality of the paired translators $(\eta,\eta')$, building on linear decoding/encoding practice \citep{yamins2014,horikawa2017} and the linear-representation hypothesis \citep{mikolov2013,park2024} as an operational assumption for the chosen $F_W$ spaces; (ii) we define NVS, a permutation-normalized residual of that naturality square, and use a cospan reading to situate existing scalar metrics by what they retain and discard; (iii) a synthetic proof-of-concept demonstrates that NVS recovers complementary world-axis sharing that standard scalars collapse; (iv) on the GOD dataset (5 subjects, 3 vision DNNs, 3 World Model proxies) we report a hierarchy crossover in which alignment is selective and axis-dependent, with animacy as the strongest among the CAV-parameterized candidate morphism families tested.

\section{Related Work}
\label{sec:related}
Most prior brain--DNN alignment work has asked whether the same \emph{objects} or \emph{stimulus sets} are represented similarly. Our framing is complementary: whether the same \emph{transformations} or \emph{relations} are preserved across the two systems.
\paragraph{Encoding/decoding metrics.} Encoding \citep{yamins2014,naselaris2011} and decoding \citep{horikawa2017} accuracies provide per-stimulus correspondence between DNNs and brains; Brain-Score \citep{schrimpf2018} introduced these as a composite benchmark and was later developed into an integrative benchmarking framework \citep{schrimpf2020}, while the BH score \citep{nonaka2021} summarizes the resulting ROI--layer hierarchy. In our framework these correspond to the closest object-level special cases of the cospan, obtained by choosing $W{=}M$ or $W{=}B$ and collapsing the question to stimuli rather than directed morphisms (App.~\ref{app:reductions}).
\paragraph{Representational similarity.} RSA \citep{kriegeskorte2008,kriegeskorte2015}, CKA \citep{kornblith2019}, and Procrustes alignment compare representational geometry through second-order pairwise statistics or global alignment transforms. These methods move from individual objects to stimulus-set structure, but still summarize agreement as a single scalar and do not isolate which candidate transformations are preserved.
\paragraph{Critical re-evaluation of brain--DNN alignment.} Recent work \citep{conwell2024} questions how strongly current scalar metrics actually distinguish brain-aligned from non-aligned representations, motivating finer-resolution diagnostics. \citet{sucholutsky2023} similarly argue alignment is plural; our proposal isolates preserved \emph{morphism families} as one such aspect, complementary to integrative-benchmark approaches \citep{schrimpf2020}. The aim here is not merely a finer scalar, but a shift in question: from global similarity to selective correspondence across morphism classes. Animacy is a long-standing organizing principle of the ventral stream \citep{konkle2013,long2018}, providing prior validation for the axis-level outlier we report.
\paragraph{World models in neuro-AI.} \citet{ha2018,bisk2020} situate a world model as a latent representation through which perception is interpreted. We use $F_W$ as a linearized proxy for this role: not as another target representation to match, but as a structural anchor that specifies which candidate morphisms are being tested.
\paragraph{Categorical / equivariant approaches.} Categorical deep learning \citep{gavranovic2024} and category-theoretic work in cognitive science \citep{ehresmann2007,phillips2022frontiers} provide formal language for structure-preserving mappings, but have not been applied to brain--DNN data. More broadly, treating brain--DNN alignment as preservation of selected relations under an externally supplied comparison map parallels the partial-structures view of scientific representation \citep{daCostaFrench2003}, in which inter-system correspondence holds only on a designated subset of relations, and the comparison-map-relative reading of representation due to \citet{vanFraassen2008}. Equivariant networks \citep{cohen2016,sanborn2023} build transformation-respecting structure directly into $F_M$ at design time, fixing which transformations a model \emph{must} preserve. Our framework is complementary: $F_M$ is taken as given (a pre-trained vision DNN), and we measure post hoc which transformations brain and DNN \emph{do} preserve in common, with the candidate class set by $F_W$ rather than baked in.

\section{Method: cospan and NVS}
\label{sec:method}

\paragraph{Spaces and estimated maps.}
Our goal is to test whether the \emph{same candidate stimulus morphism} is preserved across two systems --- a brain space $B$ (synthetic factor projection in \S\ref{sec:poc}, fMRI voxel space in \S\ref{sec:results}) and a model space $M$ (toy DNN candidates in \S\ref{sec:poc}, vision-DNN layer activations in \S\ref{sec:results}) --- relative to an external proxy space $W$ with embedding function $F_W\colon\mathbf{Stim}\to W$, so each candidate morphism is parameterized by $\Delta_W = F_W(s') - F_W(s)\in W$. The naturality square (Eq.~\ref{eq:naturality}) is realized operationally through three learned linear maps:
\begin{sloppypar}\begin{itemize}
\setlength{\itemsep}{0pt}
\item $\Phi_B\colon W\to B$ and $\Phi_M\colon W\to M$, world-to-brain and world-to-model maps fit by per-target Ridge (App.~\ref{app:methods}); these realize each tested change as the increments $\Phi_B(\Delta_W)$ in $B$ and $\Phi_M(\Delta_W)$ in $M$.
\item $\eta\colon B\to M$ and $\eta'\colon M\to B$, linear translators between the two systems, fit on per-stimulus pairs.
\end{itemize}\end{sloppypar}
The proxy embedding $F_W$ parameterizes the candidate morphism family; choosing a proxy space determines which kinds of stimulus changes can be tested. Conceptually, $F_B$ and $F_M$ act as constant-on-objects \emph{quiver representations} $\mathbf{Stim}\to\mathbf{Vect}$ sending each tested $r$ to a linear operator inside $B$ or $M$ --- not strict functors, since identities and composition across stimulus pairs are not enforced --- and $\eta, \eta'$ play the role of \emph{approximate intertwiners}: under the strict-functorial reading they would be natural transformations, but here that requirement is relaxed. Operationally, $F_B(r), F_M(r)$ are realized at first order via the proxy chain $F_W, \Phi_B, \Phi_M$ above (full categorical setup in App.~\ref{app:cat}).

\begin{figure}[!t]
  \centering
  \includegraphics[width=0.92\linewidth]{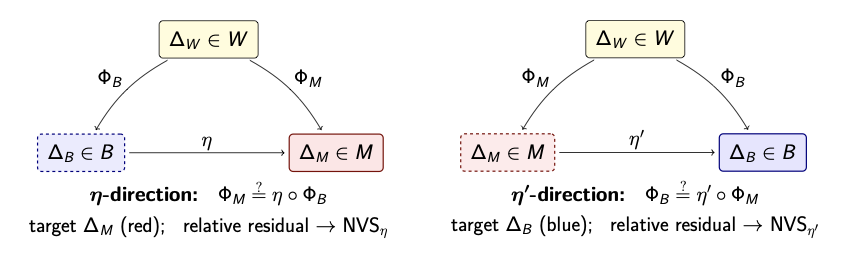}
  \caption{\textbf{Dual-triangle decomposition of NVS.} Each world morphism $\Delta_W$ has two paths to a target space. \emph{Left:} direct $\Phi_M(\Delta_W)$ vs.\ brain-mediated $\eta(\Phi_B(\Delta_W))$. \emph{Right:} direct $\Phi_B(\Delta_W)$ vs.\ DNN-mediated $\eta'(\Phi_M(\Delta_W))$. NVS is the symmetric mean of the two per-direction permutation-normalized residuals (\S\ref{sec:method}); used identically in the synthetic PoC and the empirical fMRI study.}
  \label{fig:nvs-triangle}
\end{figure}

\paragraph{NVS.}
The naturality square then becomes the operational identity
\begin{equation}
\Phi_M(\Delta_W) \;=\; \eta(\Phi_B(\Delta_W)).
\label{eq:delta-form}
\end{equation}
We measure the relative residual of Eq.~\ref{eq:delta-form} in both translation directions, normalize each direction to its own permutation null, and average. Writing $\mathbb{E}_\pi$ for the expectation under independent shufflings $\pi$ of pair indices in $\Phi_B(\Delta_W)$ and $\Phi_M(\Delta_W)$ (which preserve each side's marginal geometry while destroying cross-space correspondence; App.~\ref{app:methods}),
\begin{gather*}
\mathrm{NVS}_\eta \;=\; \frac{\mathbb{E}\,\bigl\lVert \eta(\Phi_B(\Delta_W)) - \Phi_M(\Delta_W) \bigr\rVert}{\mathbb{E}_\pi\,\bigl\lVert \eta(\Phi_B(\Delta_W)) - \Phi_M(\Delta_W) \bigr\rVert},
\\[4pt]
\mathrm{NVS}_{\eta'} \;=\; \frac{\mathbb{E}\,\bigl\lVert \eta'(\Phi_M(\Delta_W)) - \Phi_B(\Delta_W) \bigr\rVert}{\mathbb{E}_\pi\,\bigl\lVert \eta'(\Phi_M(\Delta_W)) - \Phi_B(\Delta_W) \bigr\rVert},
\\[4pt]
\mathrm{NVS} \;=\; \frac{1}{2}\bigl(\mathrm{NVS}_\eta + \mathrm{NVS}_{\eta'}\bigr).
\end{gather*}
Expectations are over ordered test pairs $(s,s')$ (per-pair Euclidean norm, then averaged). Because each $\mathrm{NVS}_d$ is already normalized to its own permutation null, the symmetric mean $\mathrm{NVS}$ is itself a chance-referenced score: $\mathrm{NVS} = 1.0$ is chance, $\mathrm{NVS} = 0$ is perfect commutativity, and lower is better. ``Chance'' here refers specifically to the permutation null that destroys cross-space pairing while preserving the marginal geometry of $\Phi_B(\Delta_W)$ and $\Phi_M(\Delta_W)$; $\mathrm{NVS} < 1$ should therefore be read as evidence for structure relative to this null, not as an absolute proof of shared representation.

The same residual form is used in two complementary regimes that share $\Phi_B, \Phi_M, \eta, \eta'$. The \emph{full-vector} regime uses the full $\Delta_W = F_W(s')-F_W(s)$ and yields $\mathrm{NVS}^{\mathrm{full}}$, a single global score over all directions in $W$. The \emph{axis-resolved} regime replaces $\Delta_W$ by its projection $\langle \Delta_W, v_a \rangle\, v_a$ onto a unit CAV direction $v_a$ for a named concept axis $a$ (animacy, real size, luminance, $\ldots$; App.~\ref{app:axis-defs}) and yields $\mathrm{NVS}^a$ (e.g., $\mathrm{NVS}^{\mathrm{animacy}}$), localizing whether the specific world-named transformation is preserved. The same definition is used for both the synthetic PoC (\S\ref{sec:poc}) and the empirical fMRI study (\S\ref{sec:results}); only the concrete spaces $B, M, F_W$ differ.

\section{Synthetic Proof of Concept}
\label{sec:poc}

These analyses reveal structure that aggregate similarity scores obscure. NVS separates alignment by generating factor, exposing how different models track position, object identity, or neither even when scalar summaries look similar.

\paragraph{Setup.} The world has 5 independent factors $w = (x, y, \text{scale}, \theta, \text{color}) \sim \mathcal N(0, I_5)$, generating $1{,}500$ stimuli. The brain $B = wA_B \in \mathbb R^{32}$ uses a random orthogonal projection retaining all 5 factors. Four DNN candidates differ only in which factors they extract: $M_\text{full}$ keeps all 5; $M_\text{pos}$ keeps only $\{x, y\}$; $M_\text{obj}$ keeps only $\{\text{scale}, \theta, \text{color}\}$; $M_\text{random}$ destroys factor structure. $M_\text{pos}$ and $M_\text{obj}$ both perfectly extract a subset of world factors, but \emph{disjoint} subsets.

\paragraph{Scalar alignment metrics.} We benchmark five canonical alignment scalars -- encoding $r$, decoding $r$, RSA, top-1 linear CCA \citep{hardoon2004}, and Procrustes alignment ($1-$disparity with optimal orthogonal rotation; \citealp{schonemann1966}) -- by treating $F_W$ as a single space (Tab.~\ref{tab:poc}, left block). \emph{None of them distinguishes $M_\text{pos}$ from $M_\text{obj}$ in a way that localizes which factors each model misses.} CCA returns $\approx 0.99$ for $M_\text{full}$, $M_\text{pos}$, and $M_\text{obj}$ alike: the 5 world factors form a maximally correlated linear subspace inside $B$ regardless of which subset $M$ retains. Procrustes does separate the models numerically ($0.91/0.26/0.48$) but cannot say \emph{which} factors are aligned. Aggregated, the five scalars agree both models are broadly aligned with $B$.

\paragraph{Per-axis and full-$\Delta$ NVS.} Restricting $\Delta_W$ to a single axis (e.g., $\Delta_W = (\Delta x, 0, 0, 0, 0)$) yields a per-axis NVS that cleanly separates the candidate models by which factors they preserve (Tab.~\ref{tab:poc}, per-axis NVS block): $M_\text{pos}$ has NVS $\approx 0.02$ on $\{x, y\}$ but $\approx 0.57$ on $\{\text{scale}, \theta, \text{color}\}$, while $M_\text{obj}$ shows the exact inverse. The full-$\Delta$ NVS (rightmost column) gives a single global score that also separates the models ($M_\text{full} \approx 0.01$, $M_\text{pos} \approx 0.32$, $M_\text{obj} \approx 0.23$, $M_\text{random} \approx 0.57$), but it does not localize \emph{which} factors are preserved or missed; that diagnosis requires the per-axis decomposition.

\begin{table}[H]
  \caption{\textbf{PoC results: static scalars, per-axis NVS, and full-$\Delta$ NVS.} Higher is better for the five static scalars; lower is better for NVS. Static scalars collapse $M_\text{pos}$ and $M_\text{obj}$ to single-number summaries; per-axis NVS reveals which factors each model preserves vs.\ misses (bold cells $=$ preserved factors); full-$\Delta$ NVS gives a global score (last column).}
  \label{tab:poc}
  \centering\footnotesize
  \setlength{\tabcolsep}{4pt}
  \renewcommand{\arraystretch}{1.10}
  \begin{tabular}{l|ccccc|ccccc|c}
    \toprule
    & \multicolumn{5}{c|}{\textbf{Static} ($\uparrow$)} & \multicolumn{5}{c|}{\textbf{Per-axis NVS} ($\downarrow$)} & \textbf{Full-$\Delta$ NVS} ($\downarrow$)\\
    Model & enc $r$ & dec $r$ & RSA & CCA & Proc. &
      $\mathrm{NVS}_x$ & $\mathrm{NVS}_y$ & $\mathrm{NVS}_\text{scale}$ & $\mathrm{NVS}_\theta$ & $\mathrm{NVS}_\text{color}$ & $\mathrm{NVS}^{\mathrm{full}}$\\
    \midrule
    $M_\text{full}$   & $0.95$ & $0.95$ & $0.97$ & $0.99$ & $0.91$ & $0.01$ & $0.01$ & $0.01$ & $0.01$ & $0.01$ & $0.01$\\
    $M_\text{pos}$    & $0.55$ & $0.84$ & $0.57$ & $0.99$ & $0.24$ & $\mathbf{0.02}$ & $\mathbf{0.02}$ & $0.57$ & $0.57$ & $0.56$ & $0.32$\\
    $M_\text{obj}$    & $0.71$ & $0.90$ & $0.72$ & $0.99$ & $0.50$ & $0.57$ & $0.57$ & $\mathbf{0.02}$ & $\mathbf{0.02}$ & $\mathbf{0.02}$ & $0.23$\\
    $M_\text{random}$ & $-0.01$ & $-0.01$ & $0.00$ & $0.41$ & $0.00$ & $0.58$ & $0.58$ & $0.58$ & $0.59$ & $0.58$ & $0.57$\\
    \bottomrule
  \end{tabular}
\end{table}

An additional bias-robustness PoC also shows that adding session-specific additive bias to $B$ collapses RSA from $1.00$ to $0.29$ while symmetric NVS rises only $0.04\!\to\!0.06$, because $\Delta$-space largely cancels additive bias (App.~\ref{app:poc-bias}).

\section{Empirical study: brain--DNN alignment on GOD}
\label{sec:results}

We next move from the toy proof of concept to empirical data, instantiating the cospan of \S\ref{sec:method} with a brain side $B$ and a model side $M$ on the GOD dataset \citep{horikawa2017}. The GOD stimuli are natural object photographs drawn from ImageNet-linked categories. fMRI responses from 5 subjects viewing these images provide $B$ as voxel patterns over $5$ ROIs (V1, V2, V3, V4, and a higher-visual-cortex ROI HVC defined as LOC $\cup$ FFA $\cup$ PPA); each of 3 vision DNNs (AlexNet, ResNet-50, ViT-B/16) is fed the same images and provides $M$ as activations over $8$ representative layers (L1--L8). Three external World Model proxies supply $F_W$ for axis decomposition: CLIP-text from the per-image AMT captions, DINOv2 and DreamSim from the images themselves. Across the $1{,}200$ single-trial training images we fit $\Phi_B, \Phi_M, \eta, \eta'$ as defined in \S\ref{sec:method}; NVS is then evaluated on the $50$ trial-averaged $35$-trial test images, using the exhaustive $50\!\times\!49 = 2{,}450$ ordered test pairs. The full grid is $5\,\mathrm{ROI} \times 8\,\mathrm{layer} \times 3\,F_W \times 6\,\mathrm{axes}$ (plus a full-$F_W$ row); per-direction $\eta, \eta'$ values are in App.~\ref{app:bidir}. We adopt a two-phase \emph{exploratory--confirmatory} design (full protocol: App.~\ref{app:methods}): all choices underlying the headline confirmatory analyses --- modeling form, axis set, proxy-viability criterion, cell grid, and test statistics --- are fixed on Sub-01, and Sub-02--05 ($n{=}4$) are then analyzed under that frozen protocol as a held-out confirmatory cohort. Sub-01-only post hoc checks (e.g., the MLP $\Phi_B,\Phi_M$ control) are reported separately in App.~\ref{app:robust}. The setup is also diagnostic: we first fit conventional object-level encode/decode-style mappings and only then ask how well those mappings satisfy the naturality criterion, rather than optimizing them for commutativity. As in the PoC, we report both the per-axis and unrestricted full-$F_W$ results.

\subsection{Per-axis NVS along candidate morphism classes}
\label{sec:results-hierarchy}

The three World Model proxies are not equally informative for every tested axis. Held-out 5-fold CAV readout from $F_W$ is high for the semantic axes (animacy: $0.68$--$0.91$; real size: $0.57$--$0.83$), strong for DreamSim on the low/mid-level visual axes, modest for CLIP-text on curvilinearity and texture energy, and poor for DINOv2 on the low/mid-level visual axes (App.~\ref{app:proxy-viability}; Tab.~\ref{tab:cav-r2-cv}). Under the proxy-viability criterion used throughout the paper (positive held-out CAV readout, $R^2>0$), CLIP-text passes 4/6 main axes, DINOv2 2/6, and DreamSim 6/6; on the 15-axis atlas the corresponding counts are 12/15, 3/15, and 15/15 (App.~\ref{app:proxy-viability}). We treat these counts as diagnostics of the World Model proxies themselves: when a proxy-axis combination fails this screen, that proxy is not an adequate comparison space for that axis under the present setup. The downstream NVS values are nonetheless retained in the pooled summaries and figures, with the failed combinations marked explicitly rather than filtered out. We therefore treat $F_W$ as part of the scientific question rather than a nuisance-free substrate: different proxies support different candidate morphism families.

\begin{figure}[!t]
  \centering
  \includegraphics[width=0.86\linewidth]{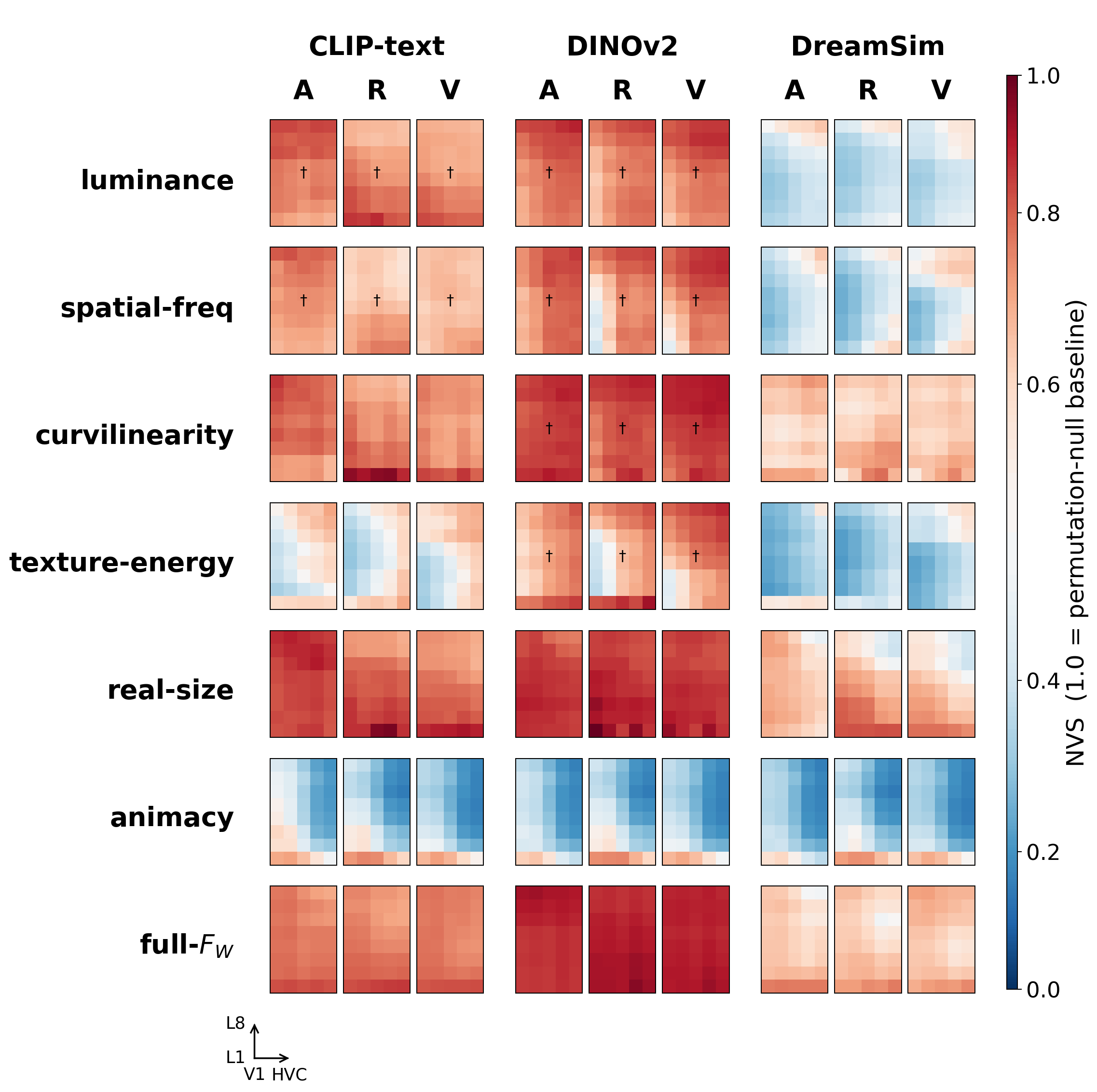}
  \caption{\textbf{5-subject mean $\mathrm{NVS}^a$ (6 scalar-axis rows) and $\mathrm{NVS}^{\mathrm{full}}$ (full-$F_W$ row, bottom) $\times$ 3 $F_W$ $\times$ 3 vision DNNs.} Each submap is a $5\,\mathrm{ROI} \times 8\,\mathrm{layer}$ heatmap, ROI horizontal (V1$\to$HVC), layer vertical (L1$\to$L8). Color: deep blue $=$ stronger preservation, white $=$ permutation null ($1.0$), red $=$ above null. \textbf{A/R/V} $=$ AlexNet/ResNet/ViT-B. The axis-resolved rows show the hierarchy crossover (low-level axes $\to$ V1$\times$shallow; semantic axes $\to$ HVC$\times$deep), absent in the diffuse full-$F_W$ row. Per-subject heatmaps: Fig.~\ref{fig:per-subject-v5}; statistical crossover: Tab.~\ref{tab:hierarchy-crossover}. Submaps failing the CAV viability screen are marked $^{\dagger}$ (in the heatmap and in the corresponding tables) and should not be read as valid tests.}
  \label{fig:8axes-heatmap}
\end{figure}

\begin{table}[H]
  \caption{\textbf{Per-axis $\mathrm{NVS}^a$ for the 6 main scalar axes: per-$F_W$ pooled mean and best cell.} The first three numeric columns give the 5-subject mean across the $3\,\mathrm{DNN}\times 5\,\mathrm{ROI}\times 8\,\mathrm{layer}$ cells for each $(F_W, \text{axis})$, reported as $\mu_{\pm\sigma}$ with $\sigma$ the across-5-subjects SD of the pooled per-subject mean. The last column gives the cell minimizing the 5-subject mean across all 9 $(F_W,\mathrm{DNN})$ blocks $\times$ $5\,\mathrm{ROI}\times 8\,\mathrm{layer}$. \textbf{A}~$=$~AlexNet, \textbf{R}~$=$~ResNet, \textbf{V}~$=$~ViT-B. Entries marked $^{\dagger}$ are proxy-axis combinations that fail the proxy-viability criterion ($R^2 \le 0$ in held-out CAV readout); they are retained for transparency and should not be interpreted as valid tests of that morphism family for that proxy.}
  \label{tab:per-condition-v5}
  \centering\footnotesize
  \setlength{\tabcolsep}{4pt}
  \renewcommand{\arraystretch}{1.10}
  \begin{tabular}{l|ccc|l}
    \toprule
                       & \multicolumn{3}{c|}{Per-$F_W$ pooled mean (over 3 DNN $\times$ 5 ROI $\times$ 8 layer)} & \\
    Axis               & CLIP-text                       & DINOv2                          & DreamSim                       & best cell (ROI$\times$L (DNN, $F_W$): $\mu_{\pm\sigma}$) \\
    \midrule
    luminance          & \nonviable{$0.680_{\pm 0.023}$} & \nonviable{$0.708_{\pm 0.036}$} & $0.448_{\pm 0.031}$            & V1$\times$L4 (R, Dream): $0.368_{\pm 0.029}$ \\
    spatial frequency  & \nonviable{$0.619_{\pm 0.073}$} & \nonviable{$0.687_{\pm 0.038}$} & $0.450_{\pm 0.059}$            & V1$\times$L4 (R, Dream): $0.320_{\pm 0.064}$ \\
    curvilinearity     & $0.691_{\pm 0.025}$             & \nonviable{$0.794_{\pm 0.017}$} & $0.579_{\pm 0.038}$            & V2$\times$L4 (A, Dream): $0.517_{\pm 0.049}$ \\
    texture energy     & $0.509_{\pm 0.074}$             & \nonviable{$0.648_{\pm 0.044}$} & $0.400_{\pm 0.042}$            & V1$\times$L2 (A, Dream): $0.278_{\pm 0.045}$ \\
    real size          & $0.749_{\pm 0.018}$             & $0.812_{\pm 0.017}$             & $0.590_{\pm 0.013}$            & HVC$\times$L7 (R, Dream): $0.448_{\pm 0.045}$ \\
    animacy            & $0.404_{\pm 0.054}$             & $0.388_{\pm 0.061}$             & $0.373_{\pm 0.063}$            & HVC$\times$L6 (R, Dream): $0.193_{\pm 0.058}$ \\
    \bottomrule
  \end{tabular}
\end{table}

Decomposing $\Delta_W$ along six scalar axes (Fig.~\ref{fig:8axes-heatmap}) reveals the main selective structure in the empirical data. The six-axis panel covers two low-level photometric axes (luminance, spatial frequency), two mid-level form/texture axes (curvilinearity, texture energy), and two higher-level semantic axes with strong ventral-stream precedent (real size, animacy). Low-level axes align best in earlier visual / shallower-layer cells and the high-level axes in higher-visual / deeper-layer cells; the two mid-level axes fall in between, with texture energy showing the cleaner intermediate pattern and curvilinearity proxy-dependent (Tab.~\ref{tab:hierarchy-crossover}). $\mathrm{NVS}^{\mathrm{real\,size}}$ pools at $0.72$ and $\mathrm{NVS}^{\mathrm{animacy}}$ at $0.39$. For animacy, this ordering is statistically clean: pooling the 5-subject mean over the 8 layers and 3 DNNs gives an ROI profile that decreases monotonically V1$\to$HVC under all three $F_W$ proxies (Spearman $\rho=-1.00$), and the layer profile likewise favors deeper layers ($\rho=-0.93, -0.93, -0.81$ for the three proxies); the same analysis on low-level axes runs in the \emph{opposite} direction (Tab.~\ref{tab:hierarchy-crossover}). The hierarchy crossover is confirmed by a within-subject permutation test on the semantic-vs-low-level class contrast of ROI-rank Spearman $\rho$ ($T=-1.34$, empirical one-sided $p<10^{-4}$, i.e.\ none of $10{,}000$ axis-label permutations reach the observed magnitude in the predicted direction, on the full 5-subject pool; the held-out confirmatory cohort Sub-02--05 ($n=4$) alone gives $T=-1.45$ with the same empirical one-sided $p<10^{-4}$; App.~\ref{app:confirmatory}). The qualitative six-axis topology is stable across subjects (Fig.~\ref{fig:per-subject-v5}, App.~\ref{app:inter-subject-v5}), and the per-axis best cells in Tab.~\ref{tab:per-condition-v5} recover an early-visual to higher-visual ROI-layer ordering analogous to BH-style results but resolved per axis \citep{nonaka2021,yamins2014}.

\subsection{Comparing axes across architectures and proxies}
\label{sec:results-animacy}

Within the tested axis set, animacy separates from the rest: 5-subject pooled $\mathrm{NVS}^{\mathrm{animacy}} = 0.39 \pm 0.06$ (bootstrap 95\% CI $[0.34, 0.44]$), the lowest-ratio axis in 5/5 subjects across the six main scalar axes (because each axis is normalized to its own permutation null, cross-axis comparisons reflect relative preservation against axis-specific shuffled baselines rather than raw residual magnitudes); across the same six axes the remaining pooled means range $0.52$--$0.72$ (Tab.~\ref{tab:per-condition-v5}). Animacy stays low across all 9 $(F_W,\mathrm{DNN})$ combinations whereas the low/mid-level axes depend more strongly on which World Model proxy is viable for them (App.~\ref{app:proxy-viability}); subject-resampling CIs and the 15-axis atlas (Apps.~\ref{app:confirmatory}, \ref{app:atlas15}) support the same axis ordering. The animate/inanimate distinction is a long-known organizing principle of ventral cortex \citep{kriegeskorte2008,konkle2013}; the morphism-level finding here is consistent with that prior without implying animacy is uniquely privileged. Crucially, ``high-level'' is not by itself sufficient for strong alignment: in the 15-axis atlas (App.~\ref{app:atlas15-with}), the affordance and material axes (\textsc{hold}, \textsc{ride}, \textsc{mat\_metal}, \textsc{mat\_natural}) all cluster at $\mathrm{NVS}^a \approx 0.73$--$0.74$, well above $\mathrm{NVS}^{\mathrm{animacy}}$ ($0.39$) and the better low/mid-level axes.

\subsection{\texorpdfstring{Unrestricted full-$F_W$ baseline}{Unrestricted full-FW baseline}}
\label{sec:results-full-fw}

Beyond the axis decomposition, the unrestricted full-$F_W$ comparison still separates the proxies (DreamSim $\approx 0.58 < $ CLIP-text $\approx 0.70 < $ DINOv2 $\approx 0.85$) but answers a different question and lacks the per-axis structure: no full-$F_W$ cell reaches the per-axis animacy minimum (Tab.~\ref{tab:fullfw-main}). The brain side $B$ and model side $M$ are the same in all three blocks; what changes is $F_W$, which specifies the morphism family being tested. Lower full-$\Delta$ NVS for DreamSim therefore means the DreamSim-defined family propagates more compatibly through the same $(B,M)$ pair than the DINOv2-defined one, i.e.\ a morphism-family comparison rather than a global proxy-quality or DNN-quality ranking; this is why \S1 treats $W$ as part of the scientific question rather than a nuisance parameter. Appendix Table~\ref{tab:fullfw-wless-app} reproduces this in the same format, and the matched W-less baseline in App.~\ref{app:wless-section} shows that the advantage is not simply a property of fitting $\eta,\eta'$ on unrestricted vectors.

\begin{table}[H]
  \caption{\textbf{Full-vector $\mathrm{NVS}^{\mathrm{full}}$ for each $(F_W,\mathrm{DNN})$ combination.} Each numeric entry is $\mu \pm \sigma$ where $\mu$ is the 5-subject mean over the relevant ROI$\times$layer cells and $\sigma$ is the across-5-subjects SD of that pooled per-subject mean. The pooled-mean column averages over all $5\,\mathrm{ROI}\times 8\,\mathrm{layer}$ cells for that $F_W$, and the best-cell column reports the single (ROI, layer, DNN) cell minimizing the 5-subject mean for that $F_W$, with $\sigma$ the across-5-subjects SD of that best cell. \textbf{A}~$=$~AlexNet, \textbf{R}~$=$~ResNet, \textbf{V}~$=$~ViT-B. This table gives the unrestricted whole-space comparison as background to the per-axis analyses above. (NVS reading convention: lower is better, $1.0$ is chance; \S\ref{app:methods} \emph{Evaluation}.)}
  \label{tab:fullfw-main}
  \centering\footnotesize
  \setlength{\tabcolsep}{4pt}
  \renewcommand{\arraystretch}{1.10}
  \begin{tabular}{lccccc}
    \toprule
    $F_W$ & A & R & V & pooled mean & best cell \\
    \midrule
    CLIP-text & $0.704 \pm 0.053$ & $0.696 \pm 0.041$ & $0.702 \pm 0.048$ & $0.701 \pm 0.047$ & HVC$\times$L8 (A): $0.626 \pm 0.046$ \\
    DINOv2    & $0.832 \pm 0.047$ & $0.859 \pm 0.041$ & $0.852 \pm 0.044$ & $0.847 \pm 0.044$ & V3$\times$L1 (A): $0.803 \pm 0.038$ \\
    DreamSim  & $0.579 \pm 0.052$ & $0.576 \pm 0.043$ & $0.590 \pm 0.051$ & $0.582 \pm 0.049$ & V4$\times$L6 (R): $0.496 \pm 0.045$ \\
    \bottomrule
  \end{tabular}
\end{table}

\subsection{Dissociation from encoding/decoding and similarity metrics}
\label{sec:results-not-encoding}

NVS is not a re-description of encoding/decoding accuracy or representational similarity. The variance decomposition (App.~\ref{app:predictor-decomp}) attributes only $\approx 34\%$ of NVS variance to five readout-quality covariates --- CAV CV $R^2$ for the axis target plus pairwise-identification accuracies of $\Phi_B$, $\Phi_M$, $\eta$, $\eta'$ --- and $\eta, \eta'$ together add only $\leq 0.7\%$ on top of the other three, while axis identity alone adds $\approx 34\%$ on top of $F_W$/DNN/subject controls. As expected for global geometry metrics, RSA and CKA on the same cells provide a single cell-level summary (peaking at $r_s=0.27$ / $0.46$) and cannot distinguish which axis drives alignment (App.~\ref{app:rsa-cka}); the session-bias PoC (App.~\ref{app:poc-bias}) further dissociates RSA (collapses $1.00 \to 0.29$ under additive bias) from symmetric NVS ($0.04 \to 0.06$, near-stable). Together, these results argue the main effect is not a by-product of simple readout strength or static geometry.

\paragraph{Role of $F_W$: W-less control.}
A W-less control (App.~\ref{app:wless}) replaces $F_W$-derived directions by independent CAVs $v_B, v_M$ in $B$ and $M$, and stays close to the permutation-null baseline on the tested axes: the resulting per-axis ratios are systematically higher than their with-$F_W$ counterparts (Tab.~\ref{tab:wless-results}). The reason is geometric: $\Phi_B(v_W), \Phi_M(v_W)$ derived from a shared $v_W$ test both sides against the \emph{same} world morphism, whereas independently optimized $v_B, v_M$ are not constrained to do so, and the cosine $\cos(\eta(v_B), v_M)$ is small (typically below $\sim 0.15$ on Sub-01/AlexNet across the tested axes). Choosing $F_W$ therefore controls \emph{which} morphism class is tested, not how strongly a fixed one is detected.

\section{Discussion}
\label{sec:discussion}

Our results recover an early-to-late ROI--layer ordering analogous to BH / Brain-Score-style analyses \citep{nonaka2021,yamins2014,schrimpf2018}, but resolved by morphism class: animacy gives the strongest support among the main axes, with navigable and texture energy also coherent in the 15-axis atlas. Existing metrics (encoding/decoding, Brain-Score, RSA, CKA, CCA, Procrustes) sit within the cospan picture as what they retain or discard (App.~\ref{app:reductions}); none directly tests agreement at the level of \emph{transformations}. Full-$\Delta$ and per-axis NVS are complementary: the full-vector regime gives a single global magnitude, while the per-axis regime localizes \emph{which} candidate transformations are preserved or missed (Tab.~\ref{tab:poc}, \S\ref{sec:results-hierarchy}). The W-less control (App.~\ref{app:wless-section}) confirms that alignment requires a shared world-side anchor.

Compared with the BH score \citep{nonaka2021}, NVS shows much weaker DNN dependence (variance decomposition: $\approx 0.001$ to DNN dummies, $\approx 0.34$ to axis identity), consistent with asking whether the same morphism class propagates through the cospan rather than how layer order maps onto V1$\to$HVC.

\paragraph{Limitations.} The empirical study uses one dataset ($n=5$ on GOD), three vision DNNs, and three $F_W$ proxies. $\eta, \eta'$ are fit with standard encoding/decoding objectives and NVS is applied only \emph{post hoc}; NVS-aware training and morphism-level losses \citep{cao2024} are natural extensions. Sharper tests will require brain recording experiments with controlled world-side transformations.

\paragraph{Broader impacts and conclusion.} Axis-resolved alignment can improve interpretability, but ``shared'' claims could still be overstated. NVS reframes alignment as naturality of $(\eta,\eta')$: \emph{which} transformations propagate determines \emph{where} alignment is strongest.

\section*{Acknowledgments}
This work was supported by JSPS KAKENHI Grant-in-Aid for Scientific Research (S) (Grant Number 25K24743) and by the Japan Agency for Medical Research and Development (AMED) (Grant Number 24wm0625409).

\bibliographystyle{abbrvnat}
\bibliography{refs}

@article{horikawa2017,
  author  = {Horikawa, Tomoyasu and Kamitani, Yukiyasu},
  title   = {Generic decoding of seen and imagined objects using hierarchical visual features},
  journal = {Nature Communications},
  volume  = {8},
  pages   = {15037},
  year    = {2017},
  doi     = {10.1038/ncomms15037}
}

@article{kamitani2005,
  author  = {Kamitani, Yukiyasu and Tong, Frank},
  title   = {Decoding the visual and subjective contents of the human brain},
  journal = {Nature Neuroscience},
  volume  = {8},
  number  = {5},
  pages   = {679--685},
  year    = {2005},
  doi     = {10.1038/nn1444}
}

@article{yamashita2008,
  author  = {Yamashita, Okito and Sato, Masa-aki and Yoshioka, Taku and Tong, Frank and Kamitani, Yukiyasu},
  title   = {Sparse estimation automatically selects voxels relevant for the decoding of {fMRI} activity patterns},
  journal = {NeuroImage},
  volume  = {42},
  number  = {4},
  pages   = {1414--1429},
  year    = {2008},
  doi     = {10.1016/j.neuroimage.2008.05.050}
}

@article{shirakawa2025,
  author  = {Shirakawa, Ken and Nagano, Yoshihiro and Tanaka, Misato and Aoki, Shuntaro C. and Muraki, Yusuke and Majima, Kei and Kamitani, Yukiyasu},
  title   = {Spurious reconstruction from brain activity},
  journal = {Neural Networks},
  volume  = {190},
  pages   = {107515},
  year    = {2025},
  doi     = {10.1016/j.neunet.2025.107515}
}

@article{schrimpf2018,
  author    = {Schrimpf, Martin and Kubilius, Jonas and Hong, Ha and Majaj, Najib J. and Rajalingham, Rishi and Issa, Elias B. and Kar, Kohitij and Bashivan, Pouya and Prescott-Roy, Jonathan and Geiger, Franziska and Poggio, Tomaso and DiCarlo, James J.},
  title     = {{Brain-Score}: Which artificial neural network for object recognition is most brain-like?},
  journal   = {bioRxiv},
  year      = {2018},
  doi       = {10.1101/407007}
}

@article{nonaka2021,
  author  = {Nonaka, Soma and Majima, Kei and Aoki, Shuntaro C. and Kamitani, Yukiyasu},
  title   = {Brain hierarchy score: Which deep neural networks are hierarchically brain-like?},
  journal = {iScience},
  volume  = {24},
  pages   = {103013},
  year    = {2021},
  doi     = {10.1016/j.isci.2021.103013}
}

@article{yamins2014,
  author  = {Yamins, Daniel L. K. and Hong, Ha and Cadieu, Charles F. and Solomon, Ethan A. and Seibert, Darren and DiCarlo, James J.},
  title   = {Performance-optimized hierarchical models predict neural responses in higher visual cortex},
  journal = {Proceedings of the National Academy of Sciences},
  volume  = {111},
  number  = {23},
  pages   = {8619--8624},
  year    = {2014},
  doi     = {10.1073/pnas.1403112111}
}

@article{kriegeskorte2008,
  author  = {Kriegeskorte, Nikolaus and Mur, Marieke and Bandettini, Peter},
  title   = {Representational similarity analysis -- connecting the branches of systems neuroscience},
  journal = {Frontiers in Systems Neuroscience},
  volume  = {2},
  pages   = {4},
  year    = {2008},
  doi     = {10.3389/neuro.06.004.2008}
}

@article{kriegeskorte2015,
  author  = {Kriegeskorte, Nikolaus},
  title   = {Deep neural networks: A new framework for modeling biological vision and brain information processing},
  journal = {Annual Review of Vision Science},
  volume  = {1},
  pages   = {417--446},
  year    = {2015},
  doi     = {10.1146/annurev-vision-082114-035447}
}

@inproceedings{kornblith2019,
  author        = {Kornblith, Simon and Norouzi, Mohammad and Lee, Honglak and Hinton, Geoffrey},
  title         = {Similarity of neural network representations revisited},
  booktitle     = {Proceedings of the International Conference on Machine Learning (ICML)},
  year          = {2019},
  volume        = {97},
  pages         = {3519--3529},
  publisher     = {PMLR},
  archivePrefix = {arXiv},
  eprint        = {1905.00414},
  url           = {https://proceedings.mlr.press/v97/kornblith19a.html}
}

@inproceedings{cohen2016,
  author        = {Cohen, Taco S. and Welling, Max},
  title         = {Group equivariant convolutional networks},
  booktitle     = {Proceedings of the International Conference on Machine Learning (ICML)},
  year          = {2016},
  volume        = {48},
  pages         = {2990--2999},
  publisher     = {PMLR},
  archivePrefix = {arXiv},
  eprint        = {1602.07576},
  url           = {https://proceedings.mlr.press/v48/cohenc16.html}
}

@inproceedings{sanborn2023,
  author    = {Sanborn, Sophia and Shewmake, Christian and Olshausen, Bruno and Hillar, Christopher},
  title     = {Bispectral neural networks},
  booktitle = {International Conference on Learning Representations (ICLR)},
  year      = {2023},
  url       = {https://openreview.net/forum?id=xnsg4pfKb7}
}

@inproceedings{mikolov2013,
  author    = {Mikolov, Tomas and Yih, Wen-tau and Zweig, Geoffrey},
  title     = {Linguistic regularities in continuous space word representations},
  booktitle = {Proceedings of the Conference of the North American Chapter of the Association for Computational Linguistics: Human Language Technologies (NAACL-HLT)},
  year      = {2013},
  pages     = {746--751},
  publisher = {Association for Computational Linguistics},
  url       = {https://aclanthology.org/N13-1090/}
}

@inproceedings{park2024,
  author        = {Park, Kiho and Choe, Yo Joong and Veitch, Victor},
  title         = {The linear representation hypothesis and the geometry of large language models},
  booktitle     = {Proceedings of the International Conference on Machine Learning (ICML)},
  year          = {2024},
  volume        = {235},
  pages         = {39643--39666},
  publisher     = {PMLR},
  archivePrefix = {arXiv},
  eprint        = {2311.03658},
  url           = {https://proceedings.mlr.press/v235/park24c.html}
}

@book{daCostaFrench2003,
  author    = {da Costa, Newton C. A. and French, Steven},
  title     = {Science and Partial Truth: A Unitary Approach to Models and Scientific Reasoning},
  publisher = {Oxford University Press},
  series    = {Oxford Studies in Philosophy of Science},
  year      = {2003}
}

@book{vanFraassen2008,
  author    = {van Fraassen, Bas C.},
  title     = {Scientific Representation: Paradoxes of Perspective},
  publisher = {Oxford University Press},
  year      = {2008}
}

@book{ehresmann2007,
  author    = {Ehresmann, Andr{\'e}e C. and Vanbremeersch, Jean-Paul},
  title     = {Memory Evolutive Systems: Hierarchy, Emergence, Cognition},
  publisher = {Elsevier},
  series    = {Studies in Multidisciplinarity},
  volume    = {4},
  year      = {2007}
}

@inproceedings{radford2021,
  author        = {Radford, Alec and Kim, Jong Wook and Hallacy, Chris and Ramesh, Aditya and Goh, Gabriel and Agarwal, Sandhini and Sastry, Girish and Askell, Amanda and Mishkin, Pamela and Clark, Jack and Krueger, Gretchen and Sutskever, Ilya},
  title         = {Learning transferable visual models from natural language supervision ({CLIP})},
  booktitle     = {Proceedings of the International Conference on Machine Learning (ICML)},
  year          = {2021},
  volume        = {139},
  pages         = {8748--8763},
  publisher     = {PMLR},
  archivePrefix = {arXiv},
  eprint        = {2103.00020},
  url           = {https://proceedings.mlr.press/v139/radford21a.html}
}

@article{oquab2023,
  author        = {Oquab, Maxime and Darcet, Timoth{\'e}e and Moutakanni, Th{\'e}o and Vo, Huy and Szafraniec, Marc and Khalidov, Vasil and Fernandez, Pierre and Haziza, Daniel and Massa, Francisco and El-Nouby, Alaaeldin and Assran, Mahmoud and Ballas, Nicolas and Galuba, Wojciech and Howes, Russell and Huang, Po-Yao and Li, Shang-Wen and Misra, Ishan and Rabbat, Michael and Sharma, Vasu and Synnaeve, Gabriel and Xu, Hu and J{\'e}gou, Herv{\'e} and Mairal, Julien and Labatut, Patrick and Joulin, Armand and Bojanowski, Piotr},
  title         = {{DINOv2}: Learning robust visual features without supervision},
  journal       = {Transactions on Machine Learning Research},
  year          = {2024},
  url           = {https://openreview.net/forum?id=a68SUt6zFt}
}

@inproceedings{fu2023,
  author        = {Fu, Stephanie and Tamir, Netanel and Sundaram, Shobhita and Chai, Lucy and Zhang, Richard and Dekel, Tali and Isola, Phillip},
  title         = {{DreamSim}: Learning new dimensions of human visual similarity using synthetic data},
  booktitle     = {Advances in Neural Information Processing Systems (NeurIPS)},
  year          = {2023},
  volume        = {36},
  pages         = {50742--50768},
  publisher     = {Curran Associates, Inc.},
  archivePrefix = {arXiv},
  eprint        = {2306.09344},
  url           = {https://papers.nips.cc/paper_files/paper/2023/hash/9f09f316a3eaf59d9ced5ffaefe97e0f-Abstract-Conference.html}
}

@inproceedings{kim2018,
  author        = {Kim, Been and Wattenberg, Martin and Gilmer, Justin and Cai, Carrie and Wexler, James and Viegas, Fernanda and Sayres, Rory},
  title         = {Interpretability beyond feature attribution: Quantitative testing with concept activation vectors ({TCAV})},
  booktitle     = {Proceedings of the International Conference on Machine Learning (ICML)},
  year          = {2018},
  volume        = {80},
  pages         = {2668--2677},
  publisher     = {PMLR},
  archivePrefix = {arXiv},
  eprint        = {1711.11279},
  url           = {https://proceedings.mlr.press/v80/kim18d.html}
}

@inproceedings{ha2018,
  author        = {Ha, David and Schmidhuber, J{\"u}rgen},
  title         = {Recurrent world models facilitate policy evolution},
  booktitle     = {Advances in Neural Information Processing Systems (NeurIPS)},
  year          = {2018},
  volume        = {31},
  pages         = {2450--2462},
  publisher     = {Curran Associates, Inc.},
  archivePrefix = {arXiv},
  eprint        = {1803.10122},
  url           = {https://papers.nips.cc/paper_files/paper/2018/hash/2de5d16682c3c35007e4e92982f1a2ba-Abstract.html}
}

@inproceedings{bisk2020,
  author        = {Bisk, Yonatan and Holtzman, Ari and Thomason, Jesse and Andreas, Jacob and Bengio, Yoshua and Chai, Joyce and Lapata, Mirella and Lazaridou, Angeliki and May, Jonathan and Nisnevich, Aleksandr and Pinto, Nicolas and Turian, Joseph},
  title         = {Experience grounds language},
  booktitle     = {Proceedings of the Conference on Empirical Methods in Natural Language Processing (EMNLP)},
  year          = {2020},
  pages         = {8718--8735},
  publisher     = {Association for Computational Linguistics},
  archivePrefix = {arXiv},
  eprint        = {2004.10151},
  doi           = {10.18653/v1/2020.emnlp-main.703},
  url           = {https://aclanthology.org/2020.emnlp-main.703/}
}

@article{cao2024,
  author  = {Cao, Rosa and Yamins, Daniel},
  title   = {Explanatory models in neuroscience, {Part} 1: Taking mechanistic abstraction seriously},
  journal = {Cognitive Systems Research},
  volume  = {87},
  pages   = {101244},
  year    = {2024},
  doi     = {10.1016/j.cogsys.2024.101244}
}

@article{cao2024b,
  author  = {Cao, Rosa and Yamins, Daniel},
  title   = {Explanatory models in neuroscience, {Part} 2: Functional intelligibility and the contravariance principle},
  journal = {Cognitive Systems Research},
  volume  = {85},
  pages   = {101200},
  year    = {2024},
  doi     = {10.1016/j.cogsys.2023.101200}
}

@article{konkle2013,
  author  = {Konkle, Talia and Caramazza, Alfonso},
  title   = {Tripartite organization of the ventral stream by animacy and object size},
  journal = {Journal of Neuroscience},
  volume  = {33},
  number  = {25},
  pages   = {10235--10242},
  year    = {2013},
  doi     = {10.1523/JNEUROSCI.0983-13.2013}
}

@inproceedings{gavranovic2024,
  author    = {Gavranovi{\'c}, Bruno and Lessard, Paul and Dudzik, Andrew and von Glehn, Tamara and Ara{\'u}jo, Jo{\~a}o G. M. and Veli{\v c}kovi{\'c}, Petar},
  title     = {Position: Categorical deep learning is an algebraic theory of all architectures},
  booktitle = {Proceedings of the International Conference on Machine Learning (ICML)},
  year      = {2024},
  volume    = {235},
  pages     = {15209--15241},
  publisher = {PMLR},
  url       = {https://proceedings.mlr.press/v235/gavranovic24a.html}
}

@article{phillips2022frontiers,
  author  = {Phillips, Steven},
  title   = {What is category theory to cognitive science? {Compositional} representation and comparison},
  journal = {Frontiers in Psychology},
  volume  = {13},
  pages   = {1048975},
  year    = {2022},
  doi     = {10.3389/fpsyg.2022.1048975}
}

@article{conwell2024,
  author  = {Conwell, Colin and Prince, Jacob S. and Kay, Kendrick N. and Alvarez, George A. and Konkle, Talia},
  title   = {A large-scale examination of inductive biases shaping high-level visual representation in brains and machines},
  journal = {Nature Communications},
  volume  = {15},
  pages   = {9383},
  year    = {2024},
  doi     = {10.1038/s41467-024-53147-y}
}

@article{sucholutsky2023,
  author  = {Sucholutsky, Ilia and Muttenthaler, Lukas and Weller, Adrian and Peng, Andi and Bobu, Andreea and Kim, Been and Love, Bradley C. and Grant, Erin and Groen, Iris and Achterberg, Jascha and Tenenbaum, Joshua B. and Collins, Katherine M. and Hermann, Katherine L. and Oktar, Kerem and Greff, Klaus and Hebart, Martin N. and Cloos, Nathan and Kriegeskorte, Nikolaus and Jacoby, Nori and Zhang, Qiuyi and Marjieh, Raja and Geirhos, Robert and Chen, Sherol and Kornblith, Simon and Rane, Sunayana and Konkle, Talia and O'Connell, Thomas P. and Unterthiner, Thomas and Lampinen, Andrew K. and M{\"u}ller, Klaus-Robert and Toneva, Mariya and Griffiths, Thomas L.},
  title   = {Getting aligned on representational alignment},
  journal = {Transactions on Machine Learning Research},
  year    = {2025},
  url     = {https://openreview.net/forum?id=Hiq7lUh4Yn}
}

@article{bao2020,
  author  = {Bao, Pinglei and She, Liang and McGill, Mason and Tsao, Doris Y.},
  title   = {A map of object space in primate inferotemporal cortex},
  journal = {Nature},
  volume  = {583},
  pages   = {103--108},
  year    = {2020},
  doi     = {10.1038/s41586-020-2350-5}
}

@article{bonnerEpstein2017,
  author  = {Bonner, Michael F. and Epstein, Russell A.},
  title   = {Coding of navigational affordances in the human visual system},
  journal = {Proceedings of the National Academy of Sciences},
  volume  = {114},
  number  = {18},
  pages   = {4793--4798},
  year    = {2017},
  doi     = {10.1073/pnas.1618228114}
}

@article{conway2009,
  author  = {Conway, Bevil R.},
  title   = {Color vision, cones, and color-coding in the cortex},
  journal = {The Neuroscientist},
  volume  = {15},
  number  = {3},
  pages   = {274--290},
  year    = {2009},
  doi     = {10.1177/1073858408331369}
}

@book{fellbaum1998,
  editor    = {Fellbaum, Christiane},
  title     = {WordNet: An Electronic Lexical Database},
  publisher = {MIT Press},
  address   = {Cambridge, MA},
  year      = {1998}
}

@article{greeneOliva2009,
  author  = {Greene, Michelle R. and Oliva, Aude},
  title   = {Recognition of natural scenes from global properties: Seeing the forest without representing the trees},
  journal = {Cognitive Psychology},
  volume  = {58},
  number  = {2},
  pages   = {137--176},
  year    = {2009},
  doi     = {10.1016/j.cogpsych.2008.06.001}
}

@article{konkleOliva2012,
  author  = {Konkle, Talia and Oliva, Aude},
  title   = {A real-world size organization of object responses in occipitotemporal cortex},
  journal = {Neuron},
  volume  = {74},
  number  = {6},
  pages   = {1114--1124},
  year    = {2012},
  doi     = {10.1016/j.neuron.2012.04.036}
}

@article{olivaTorralba2001,
  author  = {Oliva, Aude and Torralba, Antonio},
  title   = {Modeling the shape of the scene: A holistic representation of the spatial envelope},
  journal = {International Journal of Computer Vision},
  volume  = {42},
  number  = {3},
  pages   = {145--175},
  year    = {2001},
  doi     = {10.1023/A:1011139631724}
}

@article{olshausenField1996,
  author  = {Olshausen, Bruno A. and Field, David J.},
  title   = {Emergence of simple-cell receptive field properties by learning a sparse code for natural images},
  journal = {Nature},
  volume  = {381},
  number  = {6583},
  pages   = {607--609},
  year    = {1996},
  doi     = {10.1038/381607a0}
}

@article{portillaSimoncelli2000,
  author  = {Portilla, Javier and Simoncelli, Eero P.},
  title   = {A parametric texture model based on joint statistics of complex wavelet coefficients},
  journal = {International Journal of Computer Vision},
  volume  = {40},
  number  = {1},
  pages   = {49--71},
  year    = {2000},
  doi     = {10.1023/A:1026553619983}
}

@article{sharan2014,
  author  = {Sharan, Lavanya and Rosenholtz, Ruth and Adelson, Edward H.},
  title   = {Accuracy and speed of material categorization in real-world images},
  journal = {Journal of Vision},
  volume  = {14},
  number  = {9},
  pages   = {12},
  year    = {2014},
  doi     = {10.1167/14.9.12}
}

@article{schrimpf2020,
  author  = {Schrimpf, Martin and Kubilius, Jonas and Lee, Michael J. and Ratan Murty, N. Apurva and Ajemian, Robert and DiCarlo, James J.},
  title   = {Integrative benchmarking to advance neurally mechanistic models of human intelligence},
  journal = {Neuron},
  volume  = {108},
  number  = {3},
  pages   = {413--423},
  year    = {2020},
  doi     = {10.1016/j.neuron.2020.07.040}
}

@article{naselaris2011,
  author  = {Naselaris, Thomas and Kay, Kendrick N. and Nishimoto, Shinji and Gallant, Jack L.},
  title   = {Encoding and decoding in {fMRI}},
  journal = {NeuroImage},
  volume  = {56},
  number  = {2},
  pages   = {400--410},
  year    = {2011},
  doi     = {10.1016/j.neuroimage.2010.07.073}
}

@article{long2018,
  author  = {Long, Bria and Yu, Chen-Ping and Konkle, Talia},
  title   = {Mid-level visual features underlie the high-level categorical organization of the ventral stream},
  journal = {Proceedings of the National Academy of Sciences},
  volume  = {115},
  number  = {38},
  pages   = {E9015--E9024},
  year    = {2018},
  doi     = {10.1073/pnas.1719616115}
}

@article{hardoon2004,
  author  = {Hardoon, David R. and Szedm{\'a}k, S{\'a}ndor and Shawe-Taylor, John},
  title   = {Canonical correlation analysis: An overview with application to learning methods},
  journal = {Neural Computation},
  volume  = {16},
  number  = {12},
  pages   = {2639--2664},
  year    = {2004},
  doi     = {10.1162/0899766042321814}
}

@article{schonemann1966,
  author  = {Sch{\"o}nemann, Peter H.},
  title   = {A generalized solution of the orthogonal {Procrustes} problem},
  journal = {Psychometrika},
  volume  = {31},
  number  = {1},
  pages   = {1--10},
  year    = {1966},
  doi     = {10.1007/BF02289451}
}

@inproceedings{krizhevsky2012,
  title     = {{ImageNet} classification with deep convolutional neural networks},
  author    = {Krizhevsky, Alex and Sutskever, Ilya and Hinton, Geoffrey E.},
  booktitle = {Advances in Neural Information Processing Systems (NeurIPS)},
  year      = {2012},
  volume    = {25},
  pages     = {1097--1105},
  publisher = {Curran Associates, Inc.},
  url       = {https://papers.nips.cc/paper_files/paper/2012/hash/c399862d3b9d6b76c8436e924a68c45b-Abstract.html}
}

@inproceedings{he2016,
  title     = {Deep residual learning for image recognition},
  author    = {He, Kaiming and Zhang, Xiangyu and Ren, Shaoqing and Sun, Jian},
  booktitle = {Proceedings of the IEEE Conference on Computer Vision and Pattern Recognition (CVPR)},
  pages     = {770--778},
  year      = {2016}
}

@inproceedings{dosovitskiy2020,
  title     = {An image is worth 16x16 words: Transformers for image recognition at scale},
  author    = {Dosovitskiy, Alexey and Beyer, Lucas and Kolesnikov, Alexander and Weissenborn, Dirk and Zhai, Xiaohua and Unterthiner, Thomas and Dehghani, Mostafa and Minderer, Matthias and Heigold, Georg and Gelly, Sylvain and Uszkoreit, Jakob and Houlsby, Neil},
  booktitle = {International Conference on Learning Representations (ICLR)},
  year      = {2021}
}

\clearpage
\appendix
\renewcommand{\thesection}{A\arabic{section}}
\renewcommand{\thefigure}{A\arabic{figure}}
\renewcommand{\thetable}{A\arabic{table}}
\setcounter{section}{0}
\setcounter{figure}{0}
\setcounter{table}{0}

\section*{Appendix}
The appendix collects the formal definitions, extended empirical results, robustness checks, and reproducibility details underlying the main text.

\section{Categorical preliminaries}\label{app:cat}

\begin{figure}[H]
  \centering
  \includegraphics[width=0.36\linewidth]{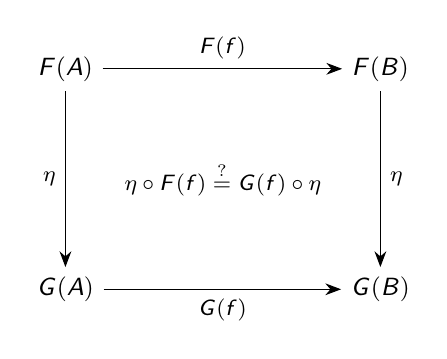}
  \caption{\textbf{Naturality square (constant-on-objects specialization).} For functors $F, G\colon\mathcal{C}\to\mathcal{D}$ that are constant on objects and a morphism $f\colon A\to B$ in $\mathcal{C}$, the natural transformation $\eta\colon F\Rightarrow G$ collapses to a single map $\eta$ which must make the square commute in $\mathcal{D}$.}
  \label{fig:nat-general}
\end{figure}

\paragraph{General categorical definitions.}
A \textbf{category} $\mathcal{C}$ consists of objects and morphisms with associative composition and identities.
A \textbf{functor} $F\colon\mathcal{C}\to\mathcal{D}$ assigns to each object of $\mathcal{C}$ an object of $\mathcal{D}$, and to each morphism a morphism, preserving composition and identities.
Given two functors $F, G\colon\mathcal{C}\to\mathcal{D}$ with the \emph{same} source and target, a \textbf{natural transformation} $\eta\colon F\Rightarrow G$ assigns to each object $A\in\mathcal{C}$ a morphism $\eta_A\colon F(A)\to G(A)$ in $\mathcal{D}$ such that the \emph{naturality square} commutes for every morphism $f\colon A\to B$ in $\mathcal{C}$. \emph{When $F, G$ are constant on objects} -- the case used throughout this paper -- all components $\eta_A$ coincide; writing $\eta$ for that single map, the naturality square becomes (Fig.~\ref{fig:nat-general}):
\begin{equation}
\eta \circ F(f) \;=\; G(f) \circ \eta.
\label{eq:nat-general}
\end{equation}
The square lives entirely in $\mathcal{D}$; the morphism $f$ in $\mathcal{C}$ enters only through its images $F(f), G(f)$.

\paragraph{Specialization to the brain--DNN setting.}
Our framework instantiates this picture as a motivational reading (the strict categorical conditions are relaxed below):
\begin{itemize}
\item $\mathcal{C} = \mathbf{Stim}$: the stimulus category. Objects are stimuli $s$; morphisms $r\colon s\to s'$ are stimulus changes.
\item $\mathcal{D} = \mathbf{Vect}$: the category of real vector spaces and linear maps. Both the brain pattern space $B$ and the DNN activation space $M$ are objects of $\mathbf{Vect}$ (with different dimensions, but in the same category).
\item $F_B\colon\mathbf{Stim}\to\mathbf{Vect}$ sends every stimulus to $B$ and each $r$ to a linear operator $F_B(r)\colon B\to B$ (Ridge-fitted from $\Delta_W \mapsto \Delta b$).
\item $F_M\colon\mathbf{Stim}\to\mathbf{Vect}$ sends every stimulus to $M$ and each $r$ to $F_M(r)\colon M\to M$.
\item Both assignments are \emph{constant on objects} ($F_B(s) = B$ and $F_M(s) = M$ for all $s$). Under the strict-categorical reading $F_B,F_M$ would be functors and the per-stimulus components of a natural transformation $\eta_s\colon F_B(s)\to F_M(s)$ would all collapse to the single linear map $\eta\colon B\to M$ -- the brain--DNN translator. We use these terms only to motivate the test; the operational version drops the functor / natural-transformation status (see \emph{Quivers and approximate functoriality} below).
\end{itemize}
The general naturality condition (Eq.~\ref{eq:nat-general}) specializes to the equation that NVS measures:
\begin{equation}
\eta \circ F_B(r) \;=\; F_M(r) \circ \eta \qquad \text{for every } r\in\mathbf{Stim}.
\label{eq:nat-ours}
\end{equation}
The corresponding square in Fig.~\ref{fig:cospan} (right panel) takes $b_s, b_{s'}, m_s, m_{s'}$ at the four corners as a concrete instantiation: $b_s$ moved by $F_B(r)$ should match $m_s$ moved by $F_M(r)$ once both are translated by $\eta$.

\paragraph{Role of the World Model.}
The World Model proxies (CLIP-text, DINOv2, DreamSim), each given by an embedding $F_W\colon \mathbf{Stim}\to W$ into a proxy space $W$, do \emph{not} appear in the naturality square. They sit above $\mathbf{Stim}$ and \emph{select} which morphisms $r$ we test, via semantic deltas $\Delta_W = F_W(s')-F_W(s)\in W$. Restricting $\Delta_W$ to a single CAV direction inside $W$ yields an axis-resolved $r$ (and hence an axis-resolved NVS). Despite the shared letter $F$, the embedding $F_W$ plays a different role from the brain- and model-side assignments $F_B,F_M$ above: $F_W$ parameterizes \emph{which} stimulus changes are tested, while $F_B,F_M$ describe how those changes are realized inside $B$ and $M$.

\paragraph{Quivers and approximate functoriality.}
The morphisms we test form a \textbf{quiver} $Q = (Q_0, Q_1)$ -- a directed multigraph with $Q_0$ stimuli and $Q_1$ edges (e.g., the semantic-delta morphisms in \S\ref{sec:method}). The operators $F_B(r), F_M(r)$ are fitted independently per edge, so we do not assume they extend to a strict functor on the free category over $Q$; composition preservation is therefore not enforced. What we fit is an \emph{empirical operator family on the quiver $Q$}, not a strict functor; ``categorical-inspired'' in our usage refers to this distinction.

\paragraph{NVS as approximate naturality.}
Empirically, Eq.~\ref{eq:nat-ours} holds only approximately. We define $\mathrm{NVS}_\eta$ (\S\ref{sec:method}) as the relative $L^2$ residual of this equation normalized to a permutation null -- a quantitative measure of approximate naturality, with both a \emph{geometric} reading (gap, in $M$, between the two paths through the square) and a \emph{statistical} one (rate of approximation failure relative to chance).

\section{PoC details}\label{app:poc-bias}

The first synthetic control corresponds to the main PoC in \S\ref{sec:poc}, where NVS separates complementary world-factor structure that standard scalar metrics collapse. The second corresponds to the brief bias-robustness remark at the end of \S\ref{sec:poc}, showing why delta-space comparisons are insensitive to additive session bias.

\subsection{Complementary-factor PoC (5-factor toy)}
World $w\in\mathbb R^5$, factors $\{x, y, \mathrm{scale}, \theta, \mathrm{color}\}\sim\mathcal N(0, I_5)$, $N=1{,}500$ stimuli. Brain $B = wA_B + \varepsilon \in \mathbb R^{32}$, $A_B$ random orthogonal. DNN candidates: $M_\text{full}$ uses $A_M$ random orthogonal; $M_\text{pos}$ masks dimensions to keep only $\{x, y\}$; $M_\text{obj}$ masks to keep $\{\mathrm{scale}, \theta, \mathrm{color}\}$; $M_\text{random}$ random projection that breaks factor structure. NVS along axis $k$ is computed with $\Delta_W = \alpha\cdot e_k$ for $\alpha\sim\mathcal N(0, 1)$, 2{,}000 samples.

\subsection{Bias-robustness PoC}
With $B = M + c_\text{session}\cdot u$ (session-specific additive bias, $|u|=1$), $c_\text{session}\sim U[-5, 5]$ across 10 sessions: object-level fit of $\eta'\colon M\to B$ degrades $1.00\to 0.66$, RSA collapses $1.00\to 0.29$, but the symmetric $\mathrm{NVS}$ rises only $0.039 \to 0.056$. The two directions decompose this asymmetrically: $\mathrm{NVS}_\eta$ is essentially unchanged ($0.039 \to 0.040$) because $\Delta$-space cancels additive bias on the $B\to M$ side, while $\mathrm{NVS}_{\eta'}$ rises modestly ($0.039 \to 0.073$) since $\eta'$ cannot recover the unobservable session-specific bias from $M$. The translator $\eta\colon B\to M$ also remains nearly intact ($1.00\to 0.999$) at the object level because the Ridge solution learns to ignore the bias direction.

\section{Detailed methods for the empirical fMRI brain--DNN analysis}\label{app:methods}

\subsection{Subjects and fMRI data}
The Generic Object Decoding (GOD) dataset \citep{horikawa2017} contains fMRI responses from 5 human subjects viewing natural object images drawn from ImageNet-linked object categories. In the standard GOD zero-shot split used here, the training set contains voxel-response patterns for $1{,}200$ training images spanning 150 object categories, each image presented once, so ``single-trial'' refers here to a single-image fMRI response pattern. The test set contains voxel-response patterns for $50$ held-out images from 50 categories that do not overlap the training categories; each test image was repeated 35 times, and the released test response for each image is the 35-trial mean fMRI pattern. Sub-01 served as the \emph{exploratory cohort}: model family, axis set, proxy-viability criterion, cell grid, and the planned hypothesis tests (e.g., sign reversal between axis classes, animacy as the lowest-ratio axis) were all fixed on Sub-01 before any inspection of Sub-02--Sub-05. Sub-02 through Sub-05 ($n{=}4$) were then held out as the \emph{confirmatory cohort}, evaluated under exactly that fixed protocol with no further model, hyperparameter, or test-statistic adjustment. This is a deliberate two-phase research-design commitment rather than a hyperparameter tuning exercise. We use 5 ROIs from the visual cortex (V1, V2, V3, V4, HVC = LOC $\cup$ FFA $\cup$ PPA).

\subsection{\texorpdfstring{World Model proxies $F_W$}{World Model proxies FW}}\label{app:fw-proxies}

\paragraph{Why call them ``World Model'' proxies?}
We follow the broad neuro-AI usage in which a ``World Model'' is a learned latent representation that encodes structural regularities of the world the agent perceives, and through which other systems can be interpreted \citep{ha2018,bisk2020}. We do not claim our $F_W$ proxies are full generative world models; rather, they are \emph{proxy embedding spaces} satisfying three structural conditions that, in our framework, are what is needed to play the role of $F_W$:
\begin{enumerate}
\setlength{\itemsep}{1pt}
\item \emph{Stimulus-aligned.} $F_W$ provides a fixed embedding $F_W(s) \in \mathbb{R}^{d_W}$ for every stimulus $s$, so $\Delta_W = F_W(s')-F_W(s)$ is well-defined as the world-side delta tested in NVS.
\item \emph{Carries world-level structure beyond the brain or the DNN.} CLIP-text encodes language-grounded semantics (captions describe categories, attributes, actions); DINOv2 encodes object-centric self-supervised image structure (instance distinctions, viewpoint invariances); DreamSim encodes human perceptual-similarity structure (judgments of which images ``look alike'' to people). Each captures a candidate organization of the world that is \emph{not} a copy of $B$ (fMRI) or any single vision DNN $M$, so $F_W$ is not collapsing into either side of the comparison.
\item \emph{Supports a meaningful axis decomposition.} Each $F_W$ admits CAV-style axis directions (App.~\ref{app:proxy-viability}; \citealp{kim2018}) that pass a held-out viability check on the candidate axes we test, ensuring that ``$\Delta_W$ along axis $k$'' is operationally well-posed.
\end{enumerate}
The viability screen (App.~\ref{app:proxy-viability}) is the empirical safeguard for condition 3: when a proxy fails on an axis, we mark it as inadequate for that morphism family rather than treating $F_W$ as a nuisance-free substrate. The three proxies were chosen to span complementary world-structure families (language, self-supervised vision, perceptual similarity) so that ``which $F_W$ supports which axis'' becomes itself part of the scientific question.

\paragraph{The three proxies used here.}
\begin{itemize}
\setlength{\itemsep}{2pt}
\item \textbf{CLIP-text} \citep{radford2021}: per-image average CLIP \emph{text} embedding from the OpenAI CLIP ViT-B/32 model (512-d). The five captions per image are AMT crowdsourced annotations supplied with the GOD stimulus set. Captures \emph{language-grounded} semantic, affordance, and material structure that humans naturally describe; weak on low-level photometric structure (captions rarely describe luminance or spatial frequency).
\item \textbf{DINOv2} \citep{oquab2023}: CLS token of ViT-B/14 self-supervised image embeddings (768-d). Captures \emph{object-centric self-supervised} image structure trained without labels on the LVD-142M corpus; strong on category-level distinctions, weak on attribute axes that captions name (its viability screen passes only animacy, real size, and navigable on the 15-axis atlas).
\item \textbf{DreamSim} \citep{fu2023}: ensemble of DINO, CLIP, and OpenCLIP backbones fine-tuned on the NIGHTS dataset of human triplet similarity judgments (512-d). Captures \emph{human perceptual similarity} structure directly and yields the lowest pooled $\mathrm{NVS}^{\mathrm{full}}$ values among the tested proxies.
\end{itemize}

\subsection{Vision DNNs}\label{app:vision-dnns}
The vision DNN side $M$ is implemented by extracting features from three architectures spanning different inductive biases. From each network we select 8 representative layers (early to late) for analysis, giving the $5\,\mathrm{ROI}\times 8\,\mathrm{layer}$ heatmap grid used throughout.
\begin{itemize}
\setlength{\itemsep}{2pt}
\item \textbf{AlexNet} \citep{krizhevsky2012}: classical 8-layer CNN trained on ImageNet-1k. We use the 5 conv stages plus the 3 fully-connected layers (cnn1--cnn8 in the heatmaps), giving a clean shallow$\to$deep progression with comparatively narrow receptive fields early on.
\item \textbf{ResNet-50} \citep{he2016}: 50-layer residual CNN trained on ImageNet-1k. We sample 8 representative blocks distributed from the stem to the last bottleneck (layer1--layer8), retaining the canonical CNN locality bias but with much greater depth and skip connections.
\item \textbf{ViT-B/16} \citep{dosovitskiy2020}: vision Transformer with 12 transformer blocks operating on $16\times 16$ patch tokens, trained on ImageNet-21k$\to$1k. We sample 8 transformer blocks (layer1--layer8) covering early to late processing. Unlike the two CNNs, ViT-B/16 has global self-attention from the first block, so the ``shallow$\to$deep'' axis is a depth axis but not a receptive-field axis.
\end{itemize}
The three architectures differ in inductive bias (locality vs.\ global attention), depth (8 vs.\ 50 vs.\ 12 blocks), and training regime, so any architecture-invariant pattern across them is a stronger claim than a pattern specific to one model family. Concretely, AlexNet and ResNet-50 use the standard \texttt{torchvision} ImageNet-1k weights (\texttt{AlexNet\_Weights.IMAGENET1K\_V1}, \texttt{ResNet50\_Weights.IMAGENET1K\_V2}); ViT-B/16 uses the \texttt{torchvision} \texttt{ViT\_B\_16\_Weights.IMAGENET1K\_V1} checkpoint, originally pretrained on ImageNet-21k and fine-tuned on ImageNet-1k.

\subsection{Per-target Ridge regression and feature selection}
\paragraph{Linear translators \texorpdfstring{$\eta,\eta'$}{eta, eta-prime}.}
The translator pair $\eta\colon B\to M$ (decoder direction) and $\eta'\colon M\to B$ (encoder direction) is modeled linearly, following the standard linear tradition in brain decoding/encoding studies \citep{kamitani2005,naselaris2011,horikawa2017,nonaka2021}. Concretely, each output dimension is fit with Ridge regression. For an output map $\phi\colon X\to Y$ and output dimension $j$, we (1) compute $|\mathrm{corr}(x_{\cdot, i}, y_{\cdot, j})|$ across training samples; (2) select the top $K=500$ input dimensions; and (3) fit Ridge regression with $\alpha=100$.

\paragraph{World-to-brain and world-to-model maps \texorpdfstring{$\Phi_B,\Phi_M$}{Phi-B, Phi-M}.}
$\Phi_B$ and $\Phi_M$ are implemented in a linear Ridge form for the main analysis, so that the maps in the empirical cospan remain first-order and directly comparable to the linear $\eta,\eta'$ layer. We additionally tested nonlinear variants of $\Phi_B,\Phi_M$ using a single-hidden-layer MLP (input $\to$ 128 ReLU units with dropout $0.3$ $\to$ output; two weight matrices, hence ``2-layer'' in the conventional MLP-counting sense), but report the linear results in the main text and treat the nonlinear results as robustness checks rather than as the primary analysis (App.~\ref{app:mlp-control}). Additional sensitivity checks for $\alpha$ and top-$K$ are reported later in the appendix.

\subsection{Semantic-delta protocol}
For training image pairs, we sample 30,000 pairs and compute $\Delta_W = F_W(s_2) - F_W(s_1)$, $\Delta_B = b_{s_2} - b_{s_1}$, and $\Delta_M = m_{s_2} - m_{s_1}$. $\Phi_B$ and $\Phi_M$ are fit from $\Delta_W \mapsto \Delta_B, \Delta_M$. Test pairs use the 50 averaged test images; we evaluate all $50\times49 = 2{,}450$ ordered pairs. By contrast, $\eta$ and $\eta'$ are fit on per-image correspondences rather than directly on deltas, following the usual per-stimulus encoding/decoding setup of \citet{naselaris2011} and later brain-decoding work. Under the present linear translator setup, however, the same fitted map also induces the corresponding delta map, since for any linear operator $L$, $L(x_2-x_1)=L(x_2)-L(x_1)$. Fitting $\eta$ and $\eta'$ at the image level therefore fixes their action on $\Delta_B$ and $\Delta_M$ as well.

\paragraph{Linear Representation Hypothesis as the rationale for \texorpdfstring{$\Delta_W$}{Delta\_W}.}
Treating $\Delta_W = F_W(s')-F_W(s)$ as a meaningful candidate stimulus morphism rests on the \emph{Linear Representation Hypothesis} (LRH; \citealp{mikolov2013,park2024}), which states that semantically meaningful features are encoded as approximately linear directions in modern representation spaces, so that vector-arithmetic differences (e.g.\ $\textsc{queen}-\textsc{woman}\approx\textsc{king}-\textsc{man}$ in the canonical word-vector example) recover interpretable transformations. Concretely, $\Delta_W$ together with the projection onto a CAV concept axis (\S\ref{app:proxy-viability}) operationalizes ``a real-world feature change'' as a linear direction in $W$. The proxy-viability check (App.~\ref{app:proxy-viability}) is the empirical safeguard for this assumption per axis: when a proxy fails to support an axis as a linear readout target (held-out CAV CV $R^2 \le 0$), we mark the corresponding $(F_W,\text{axis})$ combination as inadequate rather than rely on a non-linear direction.

\subsection{Axis definitions}\label{app:axis-defs}
We define each axis as a per-stimulus scalar target $y(s)\in\mathbb{R}$, computed from the GOD stimulus annotations or directly from the stimulus image. This subsection gives the full 15-axis appendix atlas; see \S\ref{sec:intro} and \S\ref{sec:results-hierarchy} for the six main-axis analyses.

\paragraph{Low-level photometric axes.}
\begin{itemize}\setlength{\itemsep}{2pt}
\item \textbf{luminance} (\emph{main}): mean of the per-pixel CIE luminance ($Y$ channel of CIE XYZ) over the image, after standard sRGB-to-linear conversion. Indexes overall image brightness, a parameter long known to drive early visual responses and to vary systematically across natural-image categories \citep{olivaTorralba2001}.
\item \textbf{saturation}: mean of the saturation channel (HSV $S$) over the image. Indexes how chromatic vs.\ achromatic the stimulus is, a low-level feature dissociable from hue \citep{conway2009}.
\item \textbf{hue-$a$}: mean of the CIELAB $a^*$ channel (red--green opponency).
\item \textbf{hue-$b$}: mean of the CIELAB $b^*$ channel (blue--yellow opponency). Together hue-$a$ / hue-$b$ span the perceptually approximately uniform chromatic plane used in primate cone-opponent processing \citep{conway2009}.
\end{itemize}

\paragraph{Mid-level form/texture/spatial axes.}
\begin{itemize}\setlength{\itemsep}{2pt}
\item \textbf{spatial\_freq} (\emph{main}): mean of the radially averaged Fourier-amplitude spectrum, weighted by the natural-image $1/f$ profile so that image-specific deviations from $1/f$ contribute to the scalar. Tracks the global spatial-envelope of the scene \citep{olivaTorralba2001} and aligns with the spatial frequency tuning of V1 simple cells \citep{olshausenField1996}.
\item \textbf{curvilinearity} (\emph{main}): summary statistic of orientation-tuned Gabor responses, contrasting energy at curved boundaries against energy at near-rectilinear boundaries. Curvature is a mid-level feature with dedicated tuning maps in primate inferotemporal cortex \citep{bao2020}.
\item \textbf{texture\_energy} (\emph{main}): summary scalar of multi-scale, multi-orientation Gabor-energy statistics following the texture-statistics framework of \citet{portillaSimoncelli2000}. High texture energy images carry rich repeated micro-pattern, low values are dominated by smooth surfaces.
\item \textbf{object\_area\_ratio}: ratio of segmented foreground-object area to image area, computed from per-stimulus segmentation masks supplied with the GOD annotations.
\end{itemize}

\paragraph{High-level semantic / affordance / material axes.}
\begin{itemize}\setlength{\itemsep}{2pt}
\item \textbf{animacy} (\emph{main}): binary indicator $y\in\{0,1\}$ via WordNet \citep{fellbaum1998} hypernymy, true if the stimulus's WordNet synset is a hyponym of \texttt{animal.n.01}. Animacy is a long-known organizing principle of human ventral cortex \citep{kriegeskorte2008,konkle2013,long2018}.
\item \textbf{real\_size} (\emph{main}): continuous $1$--$7$ scalar following \citet{konkleOliva2012}, assigned via a WordNet hypernym walk against a hand-curated heuristic table that maps superordinate categories (insect, mammal, vehicle, building, etc.) to a perceived real-world size rating.
\item \textbf{navigable}: binary affordance indicator, true if the stimulus depicts a region a human could plausibly walk into; constructed from the GOD per-image affordance annotations following the navigability literature \citep{greeneOliva2009,bonnerEpstein2017}.
\item \textbf{hold} / \textbf{ride}: binary affordance indicators (graspable / rideable), constructed analogously from the GOD per-image annotations.
\item \textbf{mat\_metal} / \textbf{mat\_natural}: binary material indicators (metallic surface / natural material such as wood/stone/skin), following the material-perception taxonomy of \citet{sharan2014}.
\end{itemize}

\noindent The four mid-level visual scalars (luminance, spatial frequency, curvilinearity, texture energy) and the color scalars (saturation, hue-$a$, hue-$b$) are computed per stimulus from cached per-pixel statistics released alongside the GOD stimulus archive (\texttt{code/\_god\_visual\_features\_cache.npz}, \texttt{code/\_god\_visual\_features\_extras\_cache.npz}); animacy and real size are derived from the GOD WordNet synset annotations; the affordance and material axes use the GOD per-image binary annotations. All reported axes are scalar (1-D).

\subsection{Concept-axis CAVs}
For each axis and each $F_W$, we learn a CAV $v_{\text{axis}}\in\mathbb R^{d_{F_W}}$ by Ridge regression of the axis-target scalar $y(s)$ onto $F_W$ embeddings of the same stimuli, following the concept-activation-vector formulation of \citet{kim2018}. To train $\Phi_B$ and $\Phi_M$ for a given axis, we instantiate axis-specific world deltas by sampling a scalar coefficient $\alpha\sim\mathcal{N}(0,1)$ and setting $\Delta_W=\alpha\,v_{\text{axis}}$, so the input world delta used to fit these maps varies only along that one CAV direction. For test pairs, by contrast, we start from the observed image-pair delta $\Delta_W=F_W(s')-F_W(s)$ and project it onto $v_{\text{axis}}$ to extract the component of the empirical stimulus change that lies along the chosen axis. Held-out 5-fold CV $R^2$ for axis readout from $F_W$ is reported in Tab.~\ref{tab:cav-r2-cv}.

\subsection{Proxy-viability filtering}\label{app:proxy-viability}
After defining the scalar axes, we ask which World Model proxies are adequate comparison spaces for them. For each axis and each proxy embedding $F_W$, we fit the axis readout in $W$ by 5-fold cross-validated Ridge and call the combination \emph{viable} when the held-out mean readout is positive ($R^2>0$). Failed proxy-axis combinations are retained in the main figures and pooled summaries but marked explicitly rather than dropped from downstream averaging.

\begin{table}[H]
  \caption{\textbf{Held-out 5-fold CV $R^2$ for Ridge ($F_W \to$ axis target)} for all 15 scalar axes in the appendix atlas. Entries are fold-mean $R^2$ on the $1{,}200$ training stimuli (Sub-01). Entries marked $^{\dagger}$ fail the proxy-viability criterion ($R^2 \le 0$). Across the 15-axis atlas, CLIP-text passes $12/15$ axes, DINOv2 passes $3/15$, and DreamSim passes $15/15$.}
  \label{tab:cav-r2-cv}
  \centering\footnotesize
  \setlength{\tabcolsep}{5pt}
  \renewcommand{\arraystretch}{1.08}
  \begin{tabular}{l ccc}
    \toprule
    Axis & CLIP-text & DINOv2 & DreamSim \\
    \midrule
    luminance        & \nonviable{$-0.04$} & \nonviable{$-1.24$} & 0.74 \\
    saturation       & 0.09 & \nonviable{$-1.62$} & 0.68 \\
    hue-$a$          & 0.21 & \nonviable{$-2.28$} & 0.61 \\
    hue-$b$          & 0.26 & \nonviable{$-1.86$} & 0.69 \\
    spatial\_freq    & \nonviable{$-0.06$} & \nonviable{$-1.93$} & 0.54 \\
    curvilinearity   & 0.03 & \nonviable{$-2.26$} & 0.41 \\
    texture\_energy  & 0.13 & \nonviable{$-1.11$} & 0.72 \\
    object area ratio & \nonviable{$-0.29$} & \nonviable{$-3.25$} & 0.08 \\
    real\_size       & 0.83 & 0.57 & 0.62 \\
    animacy          & 0.91 & 0.68 & 0.82 \\
    navigable        & 0.69 & 0.30 & 0.53 \\
    hold             & 0.95 & \nonviable{$-1.74$} & 0.33 \\
    ride             & 0.94 & \nonviable{$-2.09$} & 0.27 \\
    mat\_metal       & 0.95 & \nonviable{$-1.52$} & 0.37 \\
    mat\_natural     & 0.94 & \nonviable{$-2.21$} & 0.23 \\
    \bottomrule
  \end{tabular}
\end{table}

The main implication is that the three proxies should not be interpreted symmetrically. DreamSim is the most uniformly informative substrate in this study; CLIP-text is mainly useful on semantic and caption-accessible axes; DINOv2 mainly supports animacy, real size, and navigable. Outside those supported subsets, a weak result is better read as a proxy mismatch than as a decisive negative about the axis or the brain--DNN relation itself.

\subsection{Evaluation}
Evaluation is performed on the full set of ordered test-image pairs. For each pair, we compute the two path outputs in each direction and take the Euclidean residual in the corresponding target space. As in \S\ref{sec:method}, $\mathrm{NVS}_\eta$ and $\mathrm{NVS}_{\eta'}$ are the per-direction residuals normalized by their permutation-null residuals, and the reported quantity is their symmetric mean $\mathrm{NVS} = \tfrac{1}{2}(\mathrm{NVS}_\eta + \mathrm{NVS}_{\eta'})$. The permutation null shuffles pair indices independently in $\Phi_B(\Delta_W)$ and $\Phi_M(\Delta_W)$, preserving the marginal geometry on each side while destroying the tested cross-space correspondence.

\subsection{\texorpdfstring{Bidirectional aggregation of $\mathrm{NVS}_\eta, \mathrm{NVS}_{\eta'}$}{Bidirectional aggregation of NVS eta and eta-prime}}\label{app:bidir}
Averaging the two directions follows the logic of prior bidirectional brain--DNN comparisons that combine encoding- and decoding-side evidence into a single summary score, such as the hierarchy score of \citet{nonaka2021}. The two directions also vary together rather than disagreeing: across subject-averaged cells the per-cell $\mathrm{NVS}_\eta$ and $\mathrm{NVS}_{\eta'}$ values are positively correlated (Pearson $r = 0.65$, Spearman $\rho = 0.64$ on the six-axis panel; $r = 0.64$, $\rho = 0.63$ on the 15-axis atlas), so the symmetric summary is not averaging out qualitatively different signals.

\section{Cross-subject reproducibility and confirmatory tests}\label{app:primary-results}
This section reports the appendix evidence on cross-subject reproducibility, the hierarchy-crossover quantification, and the held-out confirmatory cohort.

\subsection{Inter-subject reproducibility and per-subject breakdown}\label{app:inter-subject-v5}

\begin{figure}[!t]
  \centering
  \includegraphics[width=0.92\linewidth]{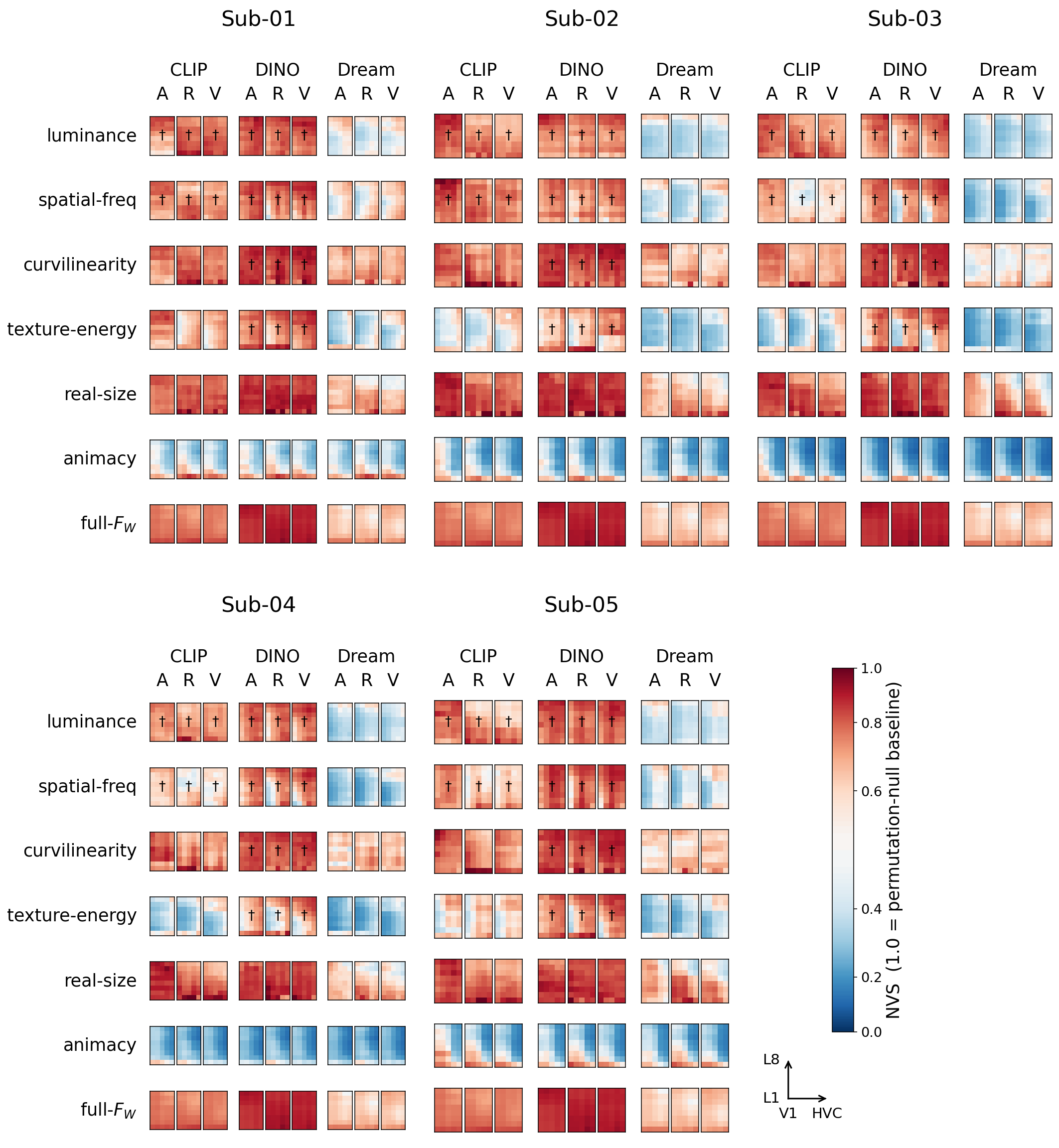}
  \caption{\textbf{Per-subject counterpart of Fig.~\ref{fig:8axes-heatmap}} (Sub-01--05, $3\!+\!2$ layout). Same axes ($\mathrm{NVS}^a$ rows + full-$F_W$ row at bottom) and same conventions (V1$\to$HVC, L1$\to$L8; A/R/V $=$ AlexNet/ResNet/ViT-B; CLIP/DINO/Dream $=$ CLIP-text/DINOv2/DreamSim). The qualitative ROI/layer structure --- low-level rows toward V1$\times$shallow, semantic rows toward HVC$\times$deep --- is stable across subjects.}
  \label{fig:per-subject-v5}
\end{figure}

The six-axis ratio profile is stable across subjects. The per-axis 5-subject pooled means and SDs (range $0.009$--$0.059$) are reported in Tab.~\ref{tab:per-condition-v5}, and the relative ordering among the six axes is preserved subject by subject. The per-subject heatmaps in Fig.~\ref{fig:per-subject-v5} show that the ROI/layer pattern is already visible within individual subjects rather than only after pooling.

\paragraph{Pattern correlation across subjects.}
For each (DNN, $F_W$, axis) submap, we compute the pairwise Pearson correlation between the (5\,ROI $\times$ 8\,layer $=$ 40-dim) ratio patterns of the 5 subjects, giving 10 pairs per submap. Across the six main scalar axes and the 9 (DNN, $F_W$) blocks (490 pairs in total), the mean $\bar r \approx 0.72$. Per-axis values span $\bar r \approx 0.50$ (curvilinearity) to $\approx 0.96$ (animacy), with texture energy $\approx 0.82$, luminance $0.74$, spatial frequency $0.70$, real size $0.57$. Per-$F_W$ stratification gives DreamSim $\bar r \approx 0.78$, DINOv2 $0.69$, CLIP-text $0.67$; per-DNN values cluster between $0.70$ and $0.74$, so cross-subject reproducibility is essentially independent of architecture and dominated by axis identity (\texttt{results/v5\_lightweight\_stats.json}, key \texttt{inter\_subject\_r}).

\subsection{Hierarchy crossover and confirmatory support}\label{app:confirmatory}

\paragraph{Hierarchy crossover: low-level vs.\ semantic axes go in opposite directions.}\label{app:hierarchy-crossover}
For each axis we compute the ROI profile (5-subject mean NVS averaged over 8 layers) and the layer profile (averaged over 5 ROIs), then take the Spearman correlation of each profile with the V1$\to$HVC and L1$\to$L8 ranks (Tab.~\ref{tab:hierarchy-crossover}). A negative $\rho$ means lower NVS toward HVC (or deeper layers); positive means lower toward V1 (or shallower). Low-level and semantic axes align in opposite directions, with mid-level axes between them.

\paragraph{Permutation test on the class contrast (5-subject pool).}
To avoid relying on the saturated rank values, we summarize each (subject, axis) by a single ROI Spearman $\rho$ (pooled across the 3 DNNs and 3 $F_W$ within the subject) and test the contrast $T = \overline{\rho}_{\mathrm{sem}} - \overline{\rho}_{\mathrm{low}}$ between the semantic class $\{\textsc{animacy}, \textsc{real\_size}\}$ and the low-level class $\{\textsc{luminance}, \textsc{spatial\_freq}\}$. Observed $T=-1.336$ on the full 5-subject pool (sem mean $-0.832$, low mean $+0.504$). Under $10{,}000$ within-subject permutations of the 6 axis labels (seed 42), $0/10{,}000$ permuted $T$ values reach the observed magnitude in the predicted direction (one-sided $p<10^{-4}$, two-sided $p\approx 10^{-4}$). The same test on the layer profile gives $T_L = -1.054$ with $0/10{,}000$ in either tail. The held-out replication on Sub-02--05 alone is reported in the next paragraph. Output: \path{results/permutation_test_axis_roi_interaction.json}.

\begin{table}[H]
  \caption{\textbf{Hierarchy crossover via Spearman $\rho$ of ROI/layer rank with $\mathrm{NVS}^a$ (5-subject mean).} For each axis $\times F_W$ proxy, we collapse the 5-subject mean NVS to (i) a 5-element ROI profile (mean over layers and 3 DNNs) and correlate with V1$\to$HVC rank; (ii) an 8-element layer profile (mean over ROIs and 3 DNNs) and correlate with L1$\to$L8 rank. Sign convention: $\rho < 0$ means NVS decreases toward HVC (resp.\ deeper layers); $\rho > 0$ means it decreases toward V1 (resp.\ shallower layers). The animacy and (DINOv2/DreamSim) real size rows show consistently negative $\rho$ on both axes; luminance, spatial frequency, and texture energy show consistently positive ROI $\rho$ for the DINOv2 and DreamSim proxies. Entries marked $^{\dagger}$ follow the convention of main-text Tab.~2. The per-subject crossover gap (Spearman $\rho_{\text{animacy}} - \rho_{\text{low-level}} \in [-2,+2]$) is reported in the surrounding text (5/5 subjects, gap $\approx -1.8$, near the $-2$ saturation bound).}
  \label{tab:hierarchy-crossover}
  \centering\footnotesize
  \setlength{\tabcolsep}{4pt}
  \begin{tabular}{l|ccc|ccc}
    \toprule
                       & \multicolumn{3}{c|}{ROI $\rho$ (V1$\to$HVC rank)} & \multicolumn{3}{c}{Layer $\rho$ (L1$\to$L8 rank)} \\
    Axis               & CLIP-text & DINOv2 & DreamSim & CLIP-text & DINOv2 & DreamSim \\
    \midrule
    luminance          & \nonviable{$-1.00$} & \nonviable{$+1.00$} & $+1.00$ & \nonviable{$-0.57$} & \nonviable{$+0.90$} & $+0.60$ \\
    spatial frequency  & \nonviable{$+0.10$} & \nonviable{$+0.90$} & $+1.00$ & \nonviable{$+0.48$} & \nonviable{$+0.98$} & $+0.43$ \\
    curvilinearity     & $-0.60$ & \nonviable{$+0.70$} & $+0.80$ & $+0.00$ & \nonviable{$+0.76$} & $-0.55$ \\
    texture energy     & $+1.00$ & \nonviable{$+1.00$} & $+1.00$ & $+0.48$ & \nonviable{$+0.50$} & $+0.26$ \\
    real size          & $+0.00$ & $-1.00$ & $-1.00$ & $-1.00$ & $-0.98$ & $-1.00$ \\
    animacy            & $-1.00$ & $-1.00$ & $-1.00$ & $-0.93$ & $-0.93$ & $-0.81$ \\
    \bottomrule
  \end{tabular}
\end{table}

\paragraph{Confirmatory cohort (Sub-02--Sub-05, $n=4$).}\label{app:resample-cellbest}
Re-running the headline statistics on the held-out cohort alone (\path{code/confirmatory_cohort_analysis.py}; output \path{results/confirmatory_cohort_summary.json}) reproduces the main pattern: (i) the per-axis pooled means preserve the full 5-subject ordering (animacy $0.362$, texture energy $0.500$, spatial frequency $0.560$, luminance $0.587$, curvilinearity $0.677$, real size $0.704$); (ii) the same class-contrast permutation test gives $T = -1.447$ on $n=4$ subjects, again $0/10{,}000$ in the predicted direction (one-sided $p < 10^{-4}$, two-sided $p \approx 10^{-4}$); (iii) the NVS variance decomposition on the $n=4$ pool ($n=7{,}840$ cells) attributes only $\approx 0.7\%$ to $\eta,\eta'$ on top of the other readout-quality covariates, while axis identity still adds $\approx +0.37$ on top of all five covariates plus $F_W$/DNN/subject controls. The held-out cohort is small ($n=4$); this analysis should be read as a within-dataset replication rather than as an independent population-level validation.

\paragraph{Bootstrap 95\% CIs (5-subject pool).}
We resample subjects with replacement ($n_\text{boot}=10{,}000$) to construct 95\% CIs on the 5-subject pooled mean ratio per axis. All five non-animacy CIs lie strictly between the animacy CI ($[0.338, 0.433]$) and the permutation-null baseline, confirming that animacy is significantly stronger than every other axis and that all six pooled main-axis means remain below that baseline. Per-axis-per-$F_W$ CIs are in \path{results/multi_subject_bootstrap_ci.json}.

\section{Atlas-wide $\mathrm{NVS}^{a}$: per-axis and full-vector baseline (15 axes)}\label{app:atlas15}\label{app:atlas15-with}

The 15-axis atlas is defined in App.~\ref{app:axis-defs}, with proxy-viability $R^2$ in Tab.~\ref{tab:cav-r2-cv}. Here we report the per-axis $\mathrm{NVS}^a$ on the full atlas (Tab.~\ref{tab:wless-best}), together with the per-DNN full-vector with-$F_W$ baseline (Tab.~\ref{tab:fullfw-wless-app}, the appendix counterpart of Tab.~\ref{tab:fullfw-main}). Animacy, navigable, and texture energy sit at the low-ratio end, while material/affordance and object-area axes cluster at $0.70$--$0.74$. DINOv2 results outside its supported subset should not be over-interpreted.

\begin{table}[H]
  \caption{\textbf{Full-vector $\mathrm{NVS}^{\mathrm{full}}$: with-$F_W$ baseline (appendix counterpart of main-text Tab.~\ref{tab:fullfw-main}).} Each numeric entry is $\mu \pm \sigma$ where $\mu$ is the 5-subject mean over $5\,\mathrm{ROI}\times 8\,\mathrm{layer}$ cells and $\sigma$ is the across-5-subjects SD of that pooled per-subject mean; the best-cell column reports the single cell minimizing the 5-subject mean. \textbf{A}~$=$~AlexNet, \textbf{R}~$=$~ResNet, \textbf{V}~$=$~ViT-B. The matched W-less control is reported separately in App.~\ref{app:wless-section}.}
  \label{tab:fullfw-wless-app}
  \centering\footnotesize
  \setlength{\tabcolsep}{4pt}
  \renewcommand{\arraystretch}{1.10}
  \begin{tabular}{lccccp{3.6cm}}
    \toprule
    $F_W$ & A & R & V & pooled mean & best cell \\
    \midrule
    CLIP-text & $0.704 \pm 0.053$ & $0.696 \pm 0.041$ & $0.702 \pm 0.048$ & $0.701 \pm 0.047$ & HVC$\times$L8 (A): $0.626 \pm 0.046$ \\
    DINOv2    & $0.832 \pm 0.047$ & $0.859 \pm 0.041$ & $0.852 \pm 0.044$ & $0.847 \pm 0.044$ & V3$\times$L1 (A): $0.803 \pm 0.038$ \\
    DreamSim  & $0.579 \pm 0.052$ & $0.576 \pm 0.043$ & $0.590 \pm 0.051$ & $0.582 \pm 0.049$ & V4$\times$L6 (R): $0.496 \pm 0.045$ \\
    \bottomrule
  \end{tabular}
\end{table}

\begin{table}[H]
  \caption{\textbf{Per-axis $\mathrm{NVS}^a$ across the 15-axis atlas (with-$F_W$).} Entries are $\mu_{\pm\sigma}$ across the 5 subjects: per-$F_W$ pooled mean over $3\,\mathrm{DNN} \times 5\,\mathrm{ROI}\times 8\,\mathrm{layer}$ cells, plus the best cell over all $9\,(F_W,\mathrm{DNN})$ blocks $\times$ $5\,\mathrm{ROI}\times 8\,\mathrm{layer}$. \textbf{A}~$=$~AlexNet, \textbf{R}~$=$~ResNet, \textbf{V}~$=$~ViT-B. Entries marked $^{\dagger}$ follow the convention of main-text Tab.~2. The matched W-less control is reported in App.~\ref{app:wless-section} (Tab.~\ref{tab:wless-results}).}
  \label{tab:wless-best}
  \centering\footnotesize
  \setlength{\tabcolsep}{4pt}
  \renewcommand{\arraystretch}{1.10}
  \begin{tabular}{lccc l}
    \toprule
    Axis               & CLIP-text                     & DINOv2                        & DreamSim                       & best cell (ROI$\times$L (DNN, $F_W$): $\mu_{\pm\sigma}$) \\
    \midrule
    luminance          & \nonviable{$0.680_{\pm 0.023}$} & \nonviable{$0.708_{\pm 0.036}$} & $0.448_{\pm 0.031}$            & V1$\times$L4 (R, Dream): $0.368_{\pm 0.029}$ \\
    saturation         & $0.662_{\pm 0.016}$             & \nonviable{$0.742_{\pm 0.033}$} & $0.495_{\pm 0.044}$            & V1$\times$L2 (A, Dream): $0.420_{\pm 0.034}$ \\
    hue-$a$            & $0.666_{\pm 0.035}$             & \nonviable{$0.786_{\pm 0.024}$} & $0.582_{\pm 0.046}$            & V2$\times$L2 (A, Dream): $0.525_{\pm 0.052}$ \\
    hue-$b$            & $0.648_{\pm 0.058}$             & \nonviable{$0.754_{\pm 0.010}$} & $0.563_{\pm 0.027}$            & HVC$\times$L1 (A, Dream): $0.489_{\pm 0.046}$ \\
    spatial frequency  & \nonviable{$0.619_{\pm 0.073}$} & \nonviable{$0.687_{\pm 0.038}$} & $0.450_{\pm 0.059}$            & V1$\times$L4 (R, Dream): $0.320_{\pm 0.064}$ \\
    curvilinearity     & $0.691_{\pm 0.025}$             & \nonviable{$0.794_{\pm 0.017}$} & $0.579_{\pm 0.038}$            & V2$\times$L4 (A, Dream): $0.517_{\pm 0.049}$ \\
    texture energy     & $0.509_{\pm 0.074}$             & \nonviable{$0.648_{\pm 0.044}$} & $0.400_{\pm 0.042}$            & V1$\times$L2 (A, Dream): $0.278_{\pm 0.045}$ \\
    object area ratio  & \nonviable{$0.680_{\pm 0.011}$} & \nonviable{$0.832_{\pm 0.008}$} & $0.577_{\pm 0.021}$            & V2$\times$L2 (A, Dream): $0.504_{\pm 0.040}$ \\
    real size          & $0.749_{\pm 0.018}$             & $0.812_{\pm 0.017}$             & $0.590_{\pm 0.013}$            & HVC$\times$L7 (R, Dream): $0.448_{\pm 0.045}$ \\
    animacy            & $0.404_{\pm 0.054}$             & $0.388_{\pm 0.061}$             & $0.373_{\pm 0.063}$            & HVC$\times$L6 (R, Dream): $0.193_{\pm 0.058}$ \\
    navigable          & $0.491_{\pm 0.043}$             & $0.496_{\pm 0.038}$             & $0.465_{\pm 0.039}$            & HVC$\times$L8 (A, Dream): $0.343_{\pm 0.066}$ \\
    hold               & $0.727_{\pm 0.020}$             & \nonviable{$0.852_{\pm 0.015}$} & $0.622_{\pm 0.032}$            & HVC$\times$L7 (R, Dream): $0.491_{\pm 0.051}$ \\
    ride               & $0.725_{\pm 0.020}$             & \nonviable{$0.849_{\pm 0.016}$} & $0.633_{\pm 0.027}$            & HVC$\times$L7 (R, Dream): $0.512_{\pm 0.053}$ \\
    mat\_metal         & $0.723_{\pm 0.020}$             & \nonviable{$0.838_{\pm 0.020}$} & $0.613_{\pm 0.027}$            & HVC$\times$L7 (R, Dream): $0.460_{\pm 0.052}$ \\
    mat\_natural       & $0.720_{\pm 0.019}$             & \nonviable{$0.846_{\pm 0.014}$} & $0.639_{\pm 0.019}$            & HVC$\times$L7 (R, Dream): $0.479_{\pm 0.058}$ \\
    \bottomrule
  \end{tabular}
\end{table}

\subsection{Atlas-wide inter-subject robustness}\label{app:atlas-robust}
Re-running the inter-subject pattern correlation and the bidirectional $\eta/\eta'$ correlation on all 15 axes without viability filtering recovers the same broad hierarchy. Animacy ($\bar r=0.96$) and navigable ($\bar r=0.85$) are the two most universal patterns across subjects; texture energy ($0.82$) and luminance ($0.74$) follow. The material and affordance axes (mat\_natural, mat\_metal, hold, ride) cluster around $\bar r \approx 0.40$--$0.43$, and object area ratio is the weakest at $\bar r=0.36$. At the atlas-wide level, the bidirectional correlation remains clearly positive even without excluding failed proxy-axis combinations (Pearson $r = 0.64$, Spearman $\rho = 0.63$ over subject-averaged cells).

\section{W-less control}\label{app:wless-section}\label{app:wless-def}\label{app:wless}\label{app:wless-interp}

We replace the shared world-side direction by axis directions trained independently in $B$ and $M$ (CAVs $v_B$ on brain features per (subject, ROI), $v_M$ on DNN features per (DNN, layer); $\eta$ from the main pipeline; same 1,000-permutation null). The full-vector W-less 3-DNN pooled mean is $0.946 \pm 0.011$, far above any with-$F_W$ proxy $0.582$--$0.847$; per-axis values are systematically higher than their with-$F_W$ counterparts on every viable axis (Tab.~\ref{tab:wless-results}), and the HVC$\times$deep minima of the main heatmap are absent. $F_W$ therefore acts as a constraint forcing both sides to be tested against the same morphism class.

\begin{table}[H]
  \caption{\textbf{W-less control: per-axis $\mathrm{NVS}^a$ across the 15-axis atlas.} Entries are $\mu_{\pm\sigma}$ across 5 subjects: per-DNN pooled mean over $5\,\mathrm{ROI}\times 8\,\mathrm{layer}$ cells, plus the pooled mean across the 3 DNNs (rightmost column). \textbf{A}~$=$~AlexNet, \textbf{R}~$=$~ResNet, \textbf{V}~$=$~ViT-B. There is no $F_W$ row because $v_B, v_M$ are trained independently in $B$ and $M$. Values stay near the permutation-null baseline and are systematically higher than their with-$F_W$ counterparts in Tab.~\ref{tab:wless-best} on every viable axis.}
  \label{tab:wless-results}
  \centering\footnotesize
  \setlength{\tabcolsep}{4pt}
  \renewcommand{\arraystretch}{1.05}
  \begin{tabular}{l|ccc|c}
    \toprule
    Axis               & A                 & R                 & V                 & pooled mean \\
    \midrule
    luminance          & $0.898_{\pm 0.017}$ & $0.879_{\pm 0.024}$ & $0.879_{\pm 0.039}$ & $0.886_{\pm 0.026}$ \\
    saturation         & $0.925_{\pm 0.028}$ & $0.875_{\pm 0.046}$ & $0.887_{\pm 0.038}$ & $0.896_{\pm 0.037}$ \\
    hue-$a$            & $0.948_{\pm 0.012}$ & $0.947_{\pm 0.011}$ & $0.942_{\pm 0.007}$ & $0.946_{\pm 0.009}$ \\
    hue-$b$            & $0.936_{\pm 0.016}$ & $0.918_{\pm 0.029}$ & $0.918_{\pm 0.020}$ & $0.924_{\pm 0.021}$ \\
    spatial frequency  & $0.871_{\pm 0.022}$ & $0.767_{\pm 0.039}$ & $0.783_{\pm 0.048}$ & $0.807_{\pm 0.033}$ \\
    curvilinearity     & $0.928_{\pm 0.013}$ & $0.859_{\pm 0.028}$ & $0.865_{\pm 0.021}$ & $0.884_{\pm 0.019}$ \\
    texture energy     & $0.825_{\pm 0.034}$ & $0.746_{\pm 0.051}$ & $0.761_{\pm 0.045}$ & $0.777_{\pm 0.041}$ \\
    object area ratio  & $0.956_{\pm 0.015}$ & $0.869_{\pm 0.033}$ & $0.894_{\pm 0.029}$ & $0.906_{\pm 0.024}$ \\
    real size          & $0.940_{\pm 0.013}$ & $0.883_{\pm 0.026}$ & $0.903_{\pm 0.022}$ & $0.909_{\pm 0.020}$ \\
    animacy            & $0.864_{\pm 0.019}$ & $0.805_{\pm 0.033}$ & $0.819_{\pm 0.019}$ & $0.830_{\pm 0.023}$ \\
    navigable          & $0.958_{\pm 0.013}$ & $0.882_{\pm 0.018}$ & $0.906_{\pm 0.019}$ & $0.915_{\pm 0.016}$ \\
    hold               & $0.939_{\pm 0.024}$ & $0.854_{\pm 0.035}$ & $0.881_{\pm 0.027}$ & $0.891_{\pm 0.028}$ \\
    ride               & $0.948_{\pm 0.021}$ & $0.874_{\pm 0.028}$ & $0.897_{\pm 0.022}$ & $0.906_{\pm 0.022}$ \\
    mat\_metal         & $0.929_{\pm 0.024}$ & $0.844_{\pm 0.032}$ & $0.870_{\pm 0.026}$ & $0.881_{\pm 0.027}$ \\
    mat\_natural       & $0.961_{\pm 0.014}$ & $0.895_{\pm 0.028}$ & $0.912_{\pm 0.023}$ & $0.923_{\pm 0.021}$ \\
    \bottomrule
  \end{tabular}
\end{table}

\section{Model diagnostics: predictive accuracy of model components}\label{app:model-diag}

Before asking what NVS is explained by (\S\ref{app:predictor-decomp}), we summarize how each component of the NVS formula behaves on its own. The translator pair $(\eta,\eta')$ is examined first (\S\ref{app:eta-pw}), then the world-to-target maps $(\Phi_B,\Phi_M)$ (\S\ref{app:fw-viability}). $\eta,\eta'$ in this section are fitted on per-stimulus pairs only and are therefore independent of the World Model proxy $F_W$ and of any concept axis. Metric consistency is preserved at the level of role: NVS tables elsewhere report NVS averaged over a common cell domain, whereas this diagnostic section uses pairwise identification for every component because these maps are defined on stimuli rather than morphism residuals and must therefore be summarized by the same object-level predictive criterion.

\subsection{$\eta, \eta'$ predictive accuracy}\label{app:eta-pw}

For each subject and each (ROI, DNN, layer) cell, we fit Ridge $\eta\colon B \to M$ on the $1{,}200$ trial-averaged training stimulus pairs and evaluate $50$-way pairwise identification on the trial-averaged test stimuli (chance $=0.5$). $\eta'\colon M \to B$ is fitted symmetrically. Tab.~\ref{tab:eta-pw} summarizes the per-cell results in a Brain-Hierarchy-style format \citep{nonaka2021}: panel~(a) reports, for each (DNN architecture, layer), the brain ROI that the layer is most accurately decoded into; panel~(b) reports, for each target brain ROI, the (DNN, layer) that encodes it most accurately. Code: \path{code/compute_eta_pure_per_cell.py}; output \path{results/eta_pure_per_cell.json}.

\begin{table}[H]
  \caption{\textbf{$\eta, \eta'$ predictive accuracy on test stimuli, $F_W$- and axis-independent (Brain-Hierarchy-style summary).} Each entry is $\mu \pm \sigma$ where $\mu$ is the 5-subject mean of $50$-way pairwise identification (chance $=0.5$) at the listed best partner and $\sigma$ is the across-5-subjects SD. $\eta, \eta'$ are fitted on per-stimulus pairs; no World Model proxy or concept axis is involved. \textbf{Panel (a)} (decoding-style $\eta\colon B\to M$): for each (DNN, layer), the brain ROI that gives the best decoding, in the spirit of \citet{nonaka2021}. \textbf{Panel (b)} (encoding-style $\eta'\colon M\to B$, transposed): for each (DNN, ROI), the DNN layer that encodes that ROI best within the same architecture.}
  \label{tab:eta-pw}
  \centering\footnotesize
  \begin{minipage}[t]{0.55\linewidth}
    \centering
    \textbf{(a) $\eta\colon B \to M$: best ROI per DNN-layer}\\[2pt]
    \setlength{\tabcolsep}{4pt}
    \begin{tabular}{l l c c}
      \toprule
      DNN & Target layer & best ROI & acc (5-subj $\mu\pm\sigma$) \\
      \midrule
      \multirow{8}{*}{AlexNet}
        & L1 & V1  & $0.620 \pm 0.033$ \\
        & L2 & V1  & $0.710 \pm 0.042$ \\
        & L3 & V2  & $0.840 \pm 0.049$ \\
        & L4 & V1  & $0.874 \pm 0.044$ \\
        & L5 & V2  & $0.756 \pm 0.032$ \\
        & L6 & HVC & $0.847 \pm 0.036$ \\
        & L7 & HVC & $0.716 \pm 0.031$ \\
        & L8 & HVC & $0.832 \pm 0.029$ \\
      \midrule
      \multirow{8}{*}{ResNet}
        & L1 & V2  & $0.547 \pm 0.026$ \\
        & L2 & V2  & $0.627 \pm 0.021$ \\
        & L3 & V2  & $0.640 \pm 0.020$ \\
        & L4 & V4  & $0.668 \pm 0.026$ \\
        & L5 & V4  & $0.707 \pm 0.029$ \\
        & L6 & HVC & $0.787 \pm 0.033$ \\
        & L7 & V4  & $0.677 \pm 0.022$ \\
        & L8 & HVC & $0.707 \pm 0.030$ \\
      \midrule
      \multirow{8}{*}{ViT-B}
        & L1 & V2  & $0.561 \pm 0.017$ \\
        & L2 & V4  & $0.664 \pm 0.019$ \\
        & L3 & V4  & $0.701 \pm 0.019$ \\
        & L4 & HVC & $0.684 \pm 0.035$ \\
        & L5 & HVC & $0.748 \pm 0.044$ \\
        & L6 & HVC & $0.636 \pm 0.022$ \\
        & L7 & V3  & $0.512 \pm 0.007$ \\
        & L8 & V3  & $0.622 \pm 0.047$ \\
      \bottomrule
    \end{tabular}
  \end{minipage}\hfill
  \begin{minipage}[t]{0.42\linewidth}
    \centering
    \textbf{(b) $\eta'\colon M \to B$: best layer per (DNN, ROI)}\\[2pt]
    \setlength{\tabcolsep}{4pt}
    \begin{tabular}{l l c c}
      \toprule
      DNN & Target ROI & best L & acc ($\mu\pm\sigma$) \\
      \midrule
      \multirow{5}{*}{AlexNet}
        & V1  & L4 & $0.577 \pm 0.070$ \\
        & V2  & L6 & $0.566 \pm 0.053$ \\
        & V3  & L7 & $0.565 \pm 0.021$ \\
        & V4  & L6 & $0.592 \pm 0.046$ \\
        & HVC & L6 & $0.586 \pm 0.074$ \\
      \midrule
      \multirow{5}{*}{ResNet}
        & V1  & L3 & $0.718 \pm 0.053$ \\
        & V2  & L3 & $0.686 \pm 0.067$ \\
        & V3  & L5 & $0.681 \pm 0.039$ \\
        & V4  & L3 & $0.660 \pm 0.077$ \\
        & HVC & L3 & $0.662 \pm 0.068$ \\
      \midrule
      \multirow{5}{*}{ViT-B}
        & V1  & L3 & $0.744 \pm 0.059$ \\
        & V2  & L3 & $0.725 \pm 0.057$ \\
        & V3  & L3 & $0.711 \pm 0.063$ \\
        & V4  & L3 & $0.696 \pm 0.082$ \\
        & HVC & L3 & $0.690 \pm 0.086$ \\
      \bottomrule
    \end{tabular}
  \end{minipage}
\end{table}

\paragraph{Consistency with prior brain--DNN translation work.} $\eta'\colon M \to B$ is the direction used in encoding-style studies \citep{yamins2014,schrimpf2018,schrimpf2020}; $\eta\colon B \to M$ is the direction used in decoding studies \citep{horikawa2017} and in the linear-readout step of the BH score \citep{nonaka2021}. The Ridge fits used here are not identical to those reference setups (e.g., \citealp{yamashita2008} uses sparse-linear regression), but the qualitative low-to-low / high-to-high correspondence between DNN layer depth and ROI hierarchy is fairly clear for AlexNet's $\eta$ panel (panel~(a): early layers L1--L4 decoded best from V1/V2, deeper layers L6--L8 from HVC) and is less clean for ResNet and ViT-B, which show only a partial monotonic progression and several non-monotonic best ROIs (e.g., ViT-B L7--L8 fall back to V3). This pattern --- clearer brain-hierarchy correspondence in AlexNet than in ResNet or ViT --- is broadly consistent with the BH score results of \citet{nonaka2021}. The encoding-side panel~(b) is similarly compatible: within each architecture, mid-to-deep layers tend to provide the best encoding into ventral-stream ROIs (AlexNet L4--L7, ResNet L3--L5, ViT-B L3). We use the per-cell pairwise-identification values directly as the readout-side covariates $X_4, X_5$ in the variance decomposition (\S\ref{app:predictor-decomp}).

\subsection{\texorpdfstring{$\Phi_B, \Phi_M$ predictive accuracy}{Phi-B, Phi-M predictive accuracy}}\label{app:fw-viability}

This subsection reports the predictive accuracy of $\Phi_B$ ($F_W \to B$) and $\Phi_M$ ($F_W \to M$) on the trial-averaged test stimuli for each proxy $F_W$ (see App.~\ref{app:proxy-viability} for the held-out CAV readout). Tab.~\ref{tab:phi-predictive} gives pairwise identification accuracies on each side.

\paragraph{Predictive accuracy of $\Phi_B$ and $\Phi_M$ across all $(F_W, \text{vision DNN})$ combinations.}
To document how well each World Model proxy $F_W$ predicts the brain side ($\Phi_B\colon F_W \to B$) and each vision DNN side ($\Phi_M\colon F_W \to M$), we trained a multi-output Ridge ($\alpha=10$) on the $1{,}200$ training stimuli and evaluated object-image-level pairwise identification on the $50$ trial-averaged test stimuli (chance $=0.5$).

\begin{table}[H]
  \caption{\textbf{Pairwise identification accuracy of the World Model proxies on the brain side (a) and the DNN side (b).} Multi-output Ridge ($\alpha=10$) trained on $1{,}200$ stimuli, evaluated with pairwise identification over $50$ test stimuli (chance $=0.5$). Panel~(a): per $(F_W, \mathrm{ROI})$, $\mu \pm \sigma$ with $\sigma$ the across-5-subjects SD. Panel~(b): per $(F_W, \mathrm{DNN})$, mean over 8 layers; the DNN side is subject-invariant, so no across-subject SD applies. The table is reported as a readout diagnostic for the tested morphism families, not as a global proxy-quality ranking.}
  \label{tab:phi-predictive}
  \centering\footnotesize
  \begin{minipage}[t]{0.42\linewidth}
    \centering
    \textbf{(a) Brain side: $\Phi_B\colon F_W \to B$}\\[2pt]
    \setlength{\tabcolsep}{4pt}
    \begin{tabular}{l l c}
      \toprule
      $F_W$ & ROI & pairwise ident.\ \\
      \midrule
      \multirow{5}{*}{CLIP-text}
        & V1  & $0.580 \pm 0.034$ \\
        & V2  & $0.574 \pm 0.041$ \\
        & V3  & $0.576 \pm 0.049$ \\
        & V4  & $0.588 \pm 0.090$ \\
        & HVC & $0.645 \pm 0.118$ \\
      \midrule
      \multirow{5}{*}{DINOv2}
        & V1  & $0.562 \pm 0.048$ \\
        & V2  & $0.573 \pm 0.022$ \\
        & V3  & $0.578 \pm 0.040$ \\
        & V4  & $0.589 \pm 0.062$ \\
        & HVC & $0.567 \pm 0.059$ \\
      \midrule
      \multirow{5}{*}{DreamSim}
        & V1  & $0.724 \pm 0.045$ \\
        & V2  & $0.707 \pm 0.043$ \\
        & V3  & $0.704 \pm 0.042$ \\
        & V4  & $0.710 \pm 0.076$ \\
        & HVC & $0.747 \pm 0.105$ \\
      \bottomrule
    \end{tabular}
  \end{minipage}\hfill
  \begin{minipage}[t]{0.55\linewidth}
    \centering
    \textbf{(b) DNN side: $\Phi_M\colon F_W \to M$}\\[2pt]
    \setlength{\tabcolsep}{3pt}
    \begin{tabular}{l l c}
      \toprule
      $F_W$ & DNN & mean (8 L) \\
      \midrule
      \multirow{3}{*}{CLIP-text}
        & AlexNet  & $0.843$ \\
        & ResNet   & $0.830$ \\
        & ViT-B    & $0.813$ \\
      \midrule
      \multirow{3}{*}{DINOv2}
        & AlexNet  & $0.771$ \\
        & ResNet   & $0.845$ \\
        & ViT-B    & $0.841$ \\
      \midrule
      \multirow{3}{*}{DreamSim}
        & AlexNet  & $0.904$ \\
        & ResNet   & $0.863$ \\
        & ViT-B    & $0.850$ \\
      \bottomrule
    \end{tabular}
  \end{minipage}
\end{table}

\paragraph{Implications.} The DreamSim $>$ CLIP-text $>$ DINOv2 ordering mirrors Tab.~\ref{tab:cav-r2-cv} and the unrestricted full-$F_W$ result, but should be read relative to the morphism families each $F_W$ makes available under the present linear setup, not as a global proxy-quality ranking. DINOv2 performs poorly on many held-out CAV readouts despite being a strong general-purpose visual representation, suggesting that the relevant proxy-axis readouts do not generalize well under the present linear setup.

\subsection{Decomposing NVS variance across the formula components}\label{app:predictor-decomp}

A direct one-line test of the dissociation in \S\ref{sec:results-not-encoding} is whether the readout-style accuracies that are routinely used by encoding/decoding-based metrics \citep{yamins2014,horikawa2017,schrimpf2018,schrimpf2020,nonaka2021} can together re-derive $\mathrm{NVS}^a$. We pool $n=10{,}000$ ROI$\times$layer$\times$DNN$\times F_W\times$subject$\times$axis cells from the 5-subject grid and fit OLS with five readout-quality covariates --- CAV CV $R^2$ ($X_1$), pairwise identification of $\Phi_B$ ($X_2$), $\Phi_M$ ($X_3$), $\eta$ ($X_4$), and $\eta'$ ($X_5$) --- plus categorical dummies for $F_W$, DNN, subject, and axis. The domain mismatch across predictors is intentional rather than ad hoc: each variable is indexed on the smallest domain on which that component is actually defined, and then broadcast to the common analysis cell only for the variance-decomposition regression. $X_4, X_5$ are reported \emph{per cell} (mean over axes within each cell), since $\eta, \eta'$ are conceptually one map per cell. Code: \path{code/analyze_nvs_predictor_decomposition_v5_with_eta.py}.

We quantify the $R^2$--$\mathrm{NVS}^a$ dissociation with three complementary tests (\path{code/analyze_r2_nvs_dissociation.py}). In a partial regression on Sub-01 cells, OLS of $\mathrm{NVS}^a$ on CV $R^2$ explains only $R^2_{\text{model}} = 0.156$; adding ROI and layer rank yields $0.177$, while adding axis intercepts raises this to $0.624$. Across the defined ($F_W$, axis) pairs, the Spearman correlation between CV $R^2$ and best-cell $\mathrm{NVS}^a$ is only $-0.45$ ($p \approx 0.08$). At the animacy hero cell (HVC$\times$L6, Sub-01), replacing the learned $\eta$ by an orthogonal Procrustes map raises $\mathrm{NVS}^{\mathrm{full}}$ from $0.93$ to $0.98$. Held-out one-sided readout quality therefore explains only a limited fraction of $\mathrm{NVS}^a$; the rest comes from axis-specific geometry and the shared translator.

\begin{table}[H]
  \caption{\textbf{$\mathrm{NVS}^a$ variance decomposition.} 5-subject pool ($n=10{,}000$ cells). The five readout-quality covariates live on different domains. We list dimensions in a uniform order, writing the (ROI, layer) pair together as the brain--DNN cell and placing subject last when present: $X_1$ varies across $(F_W,\text{axis})$, $X_2$ across $(F_W,\text{ROI},\text{subject})$, $X_3$ across $(F_W,\text{DNN},\text{layer})$, and $X_4, X_5$ across $(\text{DNN},(\text{ROI},\text{layer}),\text{subject})$ — i.e., the raw per-subject, per-cell, per-DNN $\eta, \eta'$ pairwise identification values from Tab.~\ref{tab:eta-pw} with no aggregation. $\eta, \eta'$ are themselves $F_W$- and axis-independent. ``Alone'' is univariate $R^2$; ``partial $R^2$ given $X_1\!-\!X_3$'' adds each predictor (or block) on top of the three $F_W$-explicit covariates. Together $\eta, \eta'$ add only $+0.7\%$ of NVS variance over $X_1\!-\!X_3$; axis identity still adds $+34\%$ on top of all five readout-quality covariates plus $F_W$/DNN/subject controls.}
  \label{tab:predictor-decomp}
  \centering\footnotesize
  \begin{tabular}{l c c}
    \toprule
    Predictor (domain on which it varies) & Alone $R^2$ & Partial $R^2$ given $X_1\!-\!X_3$ \\
    \midrule
    \multicolumn{3}{l}{\emph{Continuous readout-quality covariates:}} \\
    \quad $X_1$: CAV CV $R^2$, per $(F_W,\text{axis})$        & $0.264$ & --- \\
    \quad $X_2$: $\Phi_B$ pairwise ident., per $(F_W,\text{ROI},\text{subject})$  & $0.168$ & --- \\
    \quad $X_3$: $\Phi_M$ pairwise ident., per $(F_W,\text{DNN},\text{layer})$ & $0.046$ & --- \\
    \quad $X_4$: $\eta$  pairwise ident., per $(\text{DNN},(\text{ROI},\text{layer}),\text{subject})$ & $0.016$ & $+0.005$ \\
    \quad $X_5$: $\eta'$ pairwise ident., per $(\text{DNN},(\text{ROI},\text{layer}),\text{subject})$ & $0.008$ & $+0.001$ \\
    \quad $X_4 + X_5$ jointly                          & $0.023$ & $+0.007$ \\
    \midrule
    \multicolumn{3}{l}{\emph{Categorical blocks:}} \\
    \quad $F_W$ dummies (3 levels)  & $0.252$ & $+0.015$ \\
    \quad DNN dummies (3 levels)    & $0.001$ & $+0.000$ \\
    \quad Subject dummies (5 levels) & $0.039$ & $+0.026$ \\
    \quad Axis dummies (6 levels)   & $0.400$ & $+0.342$ \\
    \midrule
    \multicolumn{3}{l}{\emph{Nested model $R^2$ (cumulative):}} \\
    \quad $X_1\!-\!X_3$ (readout-quality without $\eta,\eta'$) & \multicolumn{2}{c}{$0.331$} \\
    \quad $X_1\!-\!X_5$ (readout-quality with $\eta,\eta'$)    & \multicolumn{2}{c}{$0.338$ ($\Delta=+0.007$)} \\
    \quad $+\,F_W$ dummies            & \multicolumn{2}{c}{$0.352$ ($\Delta=+0.015$)} \\
    \quad $+\,$DNN dummies            & \multicolumn{2}{c}{$0.353$ ($\Delta=+0.000$)} \\
    \quad $+\,$subject dummies        & \multicolumn{2}{c}{$0.379$ ($\Delta=+0.026$)} \\
    \quad $+\,$axis dummies           & \multicolumn{2}{c}{$0.723$ ($\Delta=+0.344$)} \\
    \bottomrule
  \end{tabular}
\end{table}

As a control for ROI-wise fMRI noise, we quantify the per-ROI noise level by split-half $\Delta_B$ reliability and add it as a sixth continuous predictor in the decomposition. This noise covariate explains only $0.011$ alone and changes the strict axis partial negligibly (\path{results/nvs_predictor_decomposition_v5_noise_ceiling.json}); the main dissociation therefore does not reduce to per-ROI fMRI noise differences. The takeaway: axis identity dominates ($\Delta R^2 \approx +0.34$ on top of all five readout scores plus $F_W$/DNN/subject), so $\mathrm{NVS}^a$ is not an axis-blind re-description of standard encoding/decoding accuracy.

\paragraph{Robustness to axis set.}
Repeating the decomposition on the 15-axis atlas (Tab.~\ref{tab:wless-best}; $n=25{,}960$, 15 axis dummies) gives the same qualitative picture: axis-identity dummies add $\approx 33\%$ on top of the readout-quality block, and the conclusion is unchanged when the per-ROI fMRI noise covariate (split-half $\Delta_B$ reliability) is added as an extra continuous predictor (\path{results/nvs_predictor_decomposition_v5_15axes_noise_ceiling.json}).
\section{Robustness analyses}\label{app:robust}
This section covers checks on whether the result could be a by-product of fitting choices or static geometry rather than genuine morphism-level structure.

\subsection{Hyperparameter sensitivity ($\alpha$ and top-$K$)}\label{app:additional-robust}
We swept Ridge $\alpha\in\{10, 100, 1000\}$ and top-$K\in\{100, 500, 2000\}$ on two representative cells (HVC$\times$ResNet$\times$layer\,6 and V1$\times$AlexNet$\times$cnn2) using DreamSim as the $F_W$ proxy on Sub-01 (\path{code/analyze_alpha_topk_v5_fast.py}; output \path{results/alpha_topk_sensitivity_v5_fast.json}). Holding top-$K$ at the main-pipeline value ($K{=}500$) and varying $\alpha$, the within-cell axis ordering is preserved at both cells in all three $\alpha$ values except that the HVC cell shifts the lowest-ratio axis between animacy and texture energy at $\alpha{=}10$. Sweeping top-$K$ produces larger magnitude shifts because $K{=}2000$ approaches the full HVC voxel count ($\sim$2{,}049) and effectively disables feature selection, but the qualitative ordering at the V1 cell is preserved across all 9 $(\alpha, K)$ combinations, and the HVC cell preserves the same lowest-ratio axis in 6/9 combinations. The main-pipeline choice $(\alpha{=}100, K{=}500)$ sits in the middle of the sweep range.

\subsection{Linear Ridge vs single-hidden-layer MLP $\Phi_B, \Phi_M$}\label{app:mlp-control}
We replaced the linear Ridge $\Phi_B, \Phi_M$ in the empirical cospan by a single-hidden-layer MLP (input $\to$ 128 ReLU units with dropout $0.3$ $\to$ output, i.e.\ two weight matrices with one ReLU nonlinearity in between) while keeping $\eta$ linear. The control is run on Sub-01 with AlexNet across the same $5\,\mathrm{ROI}\times 8\,\mathrm{layer}\times 3\,F_W$ grid as the main pipeline (script \path{code/poc_god_cav_alexnet_nonlinear_wbwm.py}; output \path{code/poc_god_cav_alexnet_nonlinear_results.json}). The matched accuracy comparison is in \path{code/analyze_wbwm_nonlinear_accuracy.py} and \path{results/wbwm_nonlinear_accuracy.json}: the nonlinear $\Phi_B, \Phi_M$ raises per-feature prediction $r$ by $+0.12$--$0.14$ on average over the linear Ridge baseline. Despite that accuracy gain, $\mathrm{NVS}^a$ shifts only modestly (axis-projected animacy moves by up to $\pm 0.30$ depending on cell), and the qualitative axis ordering on both representative cells is preserved. We therefore retain the linear Ridge form in the main text as the more interpretable first-order operator family and report the MLP variant as a Sub-01 sanity check rather than as evidence that nonlinearity is irrelevant; a full 5-subject MLP rerun is left for future work.

\subsection{RSA / CKA on identical (ROI, layer) cells}\label{app:rsa-cka}
As a scalar baseline on the same cells, we computed RSA (Spearman correlation between Pearson-distance RDMs) and linear CKA between brain test patterns and DNN test features for AlexNet, ResNet-50, and ViT-B/16 across 5 subjects. 5-subject mean best cells:
\begin{itemize}\setlength{\itemsep}{1pt}
\item RSA top-3: AlexNet$\times$cnn6$\times$HVC ($r_s=0.27$), AlexNet$\times$cnn7$\times$HVC ($0.26$), and AlexNet$\times$cnn8$\times$HVC ($0.25$).
\item CKA top-3: AlexNet$\times$cnn6$\times$HVC, AlexNet$\times$cnn4$\times$V2, and AlexNet$\times$cnn4$\times$V3 (all $0.46$).
\end{itemize}
Both RSA and CKA peak in HVC$\times$mid-deep AlexNet layers, broadly consistent with the coarse HVC/deep-layer ordering recovered by axis-resolved NVS. However, neither metric identifies which morphism class (animacy vs.\ low-level visual axes vs.\ real size) drives the alignment: a single scalar per (ROI, layer, DNN) cell collapses across the axis dimension that NVS resolves. Per-subject and per-cell values are in \path{results/rsa_cka_v5_5subj.json}.

\section{Reduction of existing metrics to NVS}\label{app:reductions}

Several commonly used brain--DNN alignment metrics can be discussed within the same broad $W$--$B$--$M$ picture, but they should not all be read as \emph{exact algebraic reductions} of NVS. Some are close to object-level special cases obtained by choosing a particular comparison space; others are better viewed as partial projections of the broader alignment problem (for example, to stimulus-set geometry rather than directed morphisms). Tab.~\ref{tab:reductions} uses this deliberately modest ``cospan reading'' only to clarify what each metric retains and what it discards. The reading is closest to exact for the object-level encoding/decoding cases; for RSA, CKA, CCA, and Procrustes it is better understood as a structural analogy indicating which aspect of the broader alignment problem is retained.

{\footnotesize
\setlength{\tabcolsep}{4pt}
\begin{longtable}{>{\raggedright\arraybackslash}p{0.27\linewidth} >{\raggedright\arraybackslash}p{0.30\linewidth} >{\raggedright\arraybackslash}p{0.36\linewidth}}
\caption{\textbf{Existing alignment metrics in the cospan picture.} NVS keeps the directed $\Delta_W$ structure explicit rather than collapsing it to objectwise prediction, global geometry, or a single shared subspace.}
\label{tab:reductions}\\
\toprule
Metric & Cospan reading & What is retained / discarded \\
\midrule
\endfirsthead
\toprule
Metric & Cospan reading & What is retained / discarded \\
\midrule
\endhead
\bottomrule
\endfoot
Encoding-based \citep{yamins2014} & Closest to an exact object-level special case: take the comparison space to be $B$ itself ($W{=}B$, $\Phi_B=\mathrm{id}$) and evaluate only $\eta'\colon M\to B$ on stimuli & Retains pointwise prediction in brain space; discards directed $\Delta_W$ structure and axis specificity \\
Decoding-based \citep{horikawa2017} & Closest to an exact object-level special case: take the comparison space to be $M$ itself ($W{=}M$, $\Phi_M=\mathrm{id}$) and evaluate only $\eta\colon B\to M$ on stimuli & Retains pointwise prediction in feature space; discards directed $\Delta_W$ structure and axis specificity \\
Brain-Score \citep{schrimpf2018,schrimpf2020} & Aggregate benchmark of encoding/decoding-style readout accuracies across multiple datasets and layers & Retains an integrative readout-quality summary; discards morphism families and the axis structure of $\Delta_W$ \\
BH score \citep{nonaka2021} & Scalar summary of object-level decoding/encoding over an ROI$\times$layer grid under one fixed feature space & Retains hierarchical localization of object-level prediction; discards morphism families and within-cell axis structure \\
RSA \citep{kriegeskorte2008,kriegeskorte2015} & Quotients the stimulus set to a representational dissimilarity matrix and compares pairwise distance structure across $B$ and $M$ & Retains symmetric second-order geometry; discards the \emph{direction}, composition, and axis identity of $\Delta_W$ \\
CKA \citep{kornblith2019} & Kernel / Gram-matrix similarity on centered activations; closely related to linear RSA rather than an exact NVS specialization & Retains global similarity of stimulus-set geometry; discards directed morphisms and axis-resolved failures \\
Linear / kernel CCA \citep{hardoon2004} & Replaces explicit world-level morphisms by maximally correlated shared subspaces between $B$ and $M$ & Retains canonical shared directions; discards non-canonical directions and can miss complementary axis failures (Tab.~\ref{tab:poc}) \\
Procrustes alignment \citep{schonemann1966} & Global alignment after constraining $\eta$ to be orthogonal; no explicit $F_W$ or directed morphism family & Retains one global geometric fit; discards axis-resolved residual structure and transformation-specific failures \\
Cao contravariance \citep{cao2024b} & Uses downstream behavior as the comparison space and studies explanatory relations with a contravariant emphasis & Retains behavior-level structure; does not instantiate the empirical world-level proxy comparison across multiple $F_W$ used here \\
Equivariant architectures / alignment priors \citep{cohen2016,sanborn2023} & Architectural complement rather than metric reduction: fixes how selected transformations act in $M$ in advance & Retains a pre-specified transformation class; does not by itself measure which morphism families are jointly preserved by brain and DNN \\
\midrule
\textbf{NVS (this paper)} & $F_W$ explicit; $\Phi_B, \Phi_M$, $\eta$, $\eta'$ all learned; directed $\Delta_W$ tested per morphism family & Retains world-level directed changes and asks whether the same candidate transformation propagates through both systems \\
\end{longtable}
}

Under this interpretation, NVS contains the object-level encoding/decoding cases as near-degenerate limits of the same setup, while RSA, CKA, CCA, Procrustes, and related approaches are more cautiously interpreted as projections or complements of the broader morphism-level question rather than as literal algebraic reductions.

\paragraph{Relation to the plural-alignment view of \citet{sucholutsky2023}.}
\citet{sucholutsky2023} argue that ``representational alignment'' is not a single quantity: different measures (encoding/decoding accuracy, RSA-style geometry, behavior-matched probes, neural predictivity) probe genuinely different aspects of how two systems correspond, and that these aspects can dissociate. The cospan reading in Tab.~\ref{tab:reductions} is a concrete realization of this stance: each existing metric is a particular projection of the $W$--$B$--$M$ picture that retains some aspects (e.g., pointwise prediction, second-order geometry, shared subspaces, orthogonal alignment) and discards others (e.g., directed $\Delta_W$, axis identity, morphism class). NVS adds a missing projection to that pluralist landscape --- preservation of \emph{directed candidate morphisms} under an explicitly chosen $F_W$ --- rather than claiming to subsume the others. In particular, our dissociations in \S\ref{sec:results-not-encoding} (NVS disagreeing with $\Phi_B$ predictive accuracy, with CAV $R^2$, and with RSA/CKA on the same cells) are empirical instances of the kind of measure-dependent partial alignment that \citet{sucholutsky2023} predict: a single ROI$\times$layer cell can score well on one projection and poorly on another, so reporting a single scalar erases the structure that the morphism-level reading recovers.

\section{Reproducibility}\label{app:repro}

\paragraph{Data.} The GOD dataset \citep{horikawa2017} for all five subjects is publicly available from figshare 7387130. As documented on the official GOD/OpenData project pages, the preprocessed GOD release on figshare is distributed under CC BY 4.0, whereas the raw fMRI data are available through OpenNeuro (dataset \texttt{ds001246}; DOI \path{10.18112/openneuro.ds001246.v1.0.1}); for copyright reasons, the original stimulus images themselves are not redistributed with this paper and remain subject to the original GOD release/request terms. The AMT-crowdsourced caption annotations used by the CLIP-text proxy are released through the same GOD distribution channel: the training-stimulus captions are publicly downloadable, whereas the test-stimulus captions are shared on request rather than fully posted online, so as to minimize the chance that they are absorbed into future model-training corpora via web crawling and thereby compromise the reusability of the GOD test set as a held-out evaluation. The supplementary \path{README.md} documents how to obtain both.

The human-subject fMRI dataset reused here was originally collected under institutional ethics approval at the time of acquisition, as documented in the cited original dataset paper \citep{horikawa2017}. Participant-facing experimental procedures, including the original task instructions and compensation/consent details for the human-data and crowdsourced-annotation components, are documented in the original dataset papers and public release materials cited here rather than reproduced.

\paragraph{Licenses and checkpoints for reused models and code.} The external model families used only as fixed feature extractors are credited and used under their public release terms; the license identifiers were checked against the corresponding public repositories used to obtain code/weights. Specifically, CLIP ViT-B/32 was loaded from \texttt{openai/clip} (MIT License), DINOv2 ViT-B/14 from the Meta AI \texttt{facebookresearch/dinov2} repository (Apache License 2.0), and DreamSim from \texttt{ssundaram21/dreamsim} (MIT License); the vision DNNs (AlexNet, ResNet-50, ViT-B/16) used the standard \texttt{torchvision} ImageNet checkpoints listed in App.~\ref{app:vision-dnns}. We do not redistribute third-party model weights in the supplementary archive; the code documents how to obtain or invoke the original public releases.

\paragraph{Code and reproducibility.} A supplementary archive (\path{supplementary_v5.zip}, $\sim$6\,MB compressed) accompanies this paper and contains the full code base required to reproduce every reported number, table, and figure end-to-end once the GOD inputs are placed at \path{./god_data/}. It bundles every \path{.py} analysis and rendering script and every \path{.json} result file referenced in this paper, together with the synthetic-PoC result bundles kept under \path{code/}, the per-stimulus low/mid-level visual-feature caches (\path{code/_god_visual_features*.npz}), a \path{requirements.txt}, and a \path{README.md} / \path{MANIFEST.md} pair documenting the layout, exact commands, and the script$\to$result/figure mapping. All paths inside scripts are relative to the supplementary root. The 25\,GB of per-subject CAV-weight caches (\path{code/_god_cav_weights_cache_*}) are not bundled because they are deterministic outputs of the per-subject CAV scripts and would inflate the archive without adding reproducibility. The archive is released under the MIT license. The pipeline is organized in three layers:
\begin{itemize}
\setlength{\itemsep}{1pt}
\item \emph{Per-subject NVS computation} — \path{code/poc_god_cav_axis_per_roi_layer.py}, \path{code/poc_god_cav_alt_dnn_pipeline.py}, \path{code/poc_god_cav_all_subspaces_addon.py}, \path{code/poc_god_cav_affordance_addon.py}, and \path{code/poc_god_cav_action_axes_addon.py} (per-subject CAV/NVS across the 6-axis main panel and the 15-axis atlas); \path{code/compute_wb_wm_full_grid.py} ($\Phi_B$/$\Phi_M$ diagnostics, App.~\ref{app:fw-viability}); \path{code/compute_eta_pure_per_cell.py} ($F_W$- and axis-independent $\eta,\eta'$, App.~\ref{app:eta-pw}; output \path{results/eta_pure_per_cell.json}); \path{code/analyze_wless_full_grid_no_axis.py} and \path{code/compute_wless_v5_15axes_5subj_pooled.py} (W-less controls; outputs \path{results/wless_v5_15axes_5subj_pooled.json} and \path{results/wless_v5_15axes_5subj_summary.json}; App.~\ref{app:wless-section}).
\item \emph{Aggregation \& statistics} — \path{code/analyze_v5_lightweight_stats.py} and \path{code/analyze_v5_15axes_lightweight.py} (per-axis pooled means, inter-subject pattern correlation, paired gap CIs; outputs \path{results/v5_lightweight_stats.json} and \path{results/v5_lightweight_stats_15axes.json}); \path{code/analyze_nvs_predictor_decomposition_v5_with_eta.py} together with its helper \path{code/analyze_nvs_predictor_decomposition_v5.py} (Tab.~\ref{tab:predictor-decomp}); \path{code/compute_noise_ceiling_v5_5subj.py} (per-ROI fMRI noise level via split-half $\Delta_B$ reliability; output \path{results/noise_ceiling_v5_5subj.json}) and \path{code/analyze_noise_ceiling_controls_v5.py} / \path{code/analyze_noise_ceiling_controls_v5_15axes.py} (ceiling-adjusted summaries; outputs \path{results/noise_ceiling_nvs_stats_v5.json}, \path{results/noise_ceiling_nvs_stats_v5_15axes.json}, \path{results/nvs_predictor_decomposition_v5_noise_ceiling.json}, and \path{results/nvs_predictor_decomposition_v5_15axes_noise_ceiling.json}); \path{code/compute_rsa_cka_v5_5subj.py} (App.~\ref{app:rsa-cka}; output \path{results/rsa_cka_v5_5subj.json}); \path{code/analyze_5subject_bootstrap_ci.py} (App.~\ref{app:confirmatory}; output \path{results/multi_subject_bootstrap_ci.json}); \path{code/permutation_test_axis_roi_interaction.py} (App.~\ref{app:hierarchy-crossover}; output \path{results/permutation_test_axis_roi_interaction.json}); \path{code/confirmatory_cohort_analysis.py} (App.~\ref{app:confirmatory}; output \path{results/confirmatory_cohort_summary.json}); \path{code/poc_nvs_world_operation_profile.py} (synthetic PoC, Tab.~\ref{tab:poc}; output \path{code/poc_nvs_world_operation_profile_results.json}); \path{code/poc_nvs_seed_average.py} (deterministic seed-averaged variant of the same PoC bundle); and \path{code/poc_nvs_advantage_drift_bias.py} (session-bias robustness PoC for App.~\ref{app:poc-bias}; output \path{code/poc_nvs_advantage_drift_bias_results.json}).
\item \emph{Figure rendering} — \path{code/figure_subject_average_publication.py} (engine), with viability-masked wrappers \path{code/render_v5_main_6axes_viability_masked.py} (Fig.~\ref{fig:8axes-heatmap}) and \path{code/render_v5_per_subject_6axes_viability_masked.py} (Fig.~\ref{fig:per-subject-v5}); \path{code/render_wless_v5_sub01_6axes.py} for the deleted W-less heatmap panel.
\end{itemize}
Stimulus annotation tables and axis-label builders for the 6 main and 15 appendix scalar axes are shipped alongside, so each axis can be regenerated from the raw GOD-side metadata.

\paragraph{Computational resources.} The pipeline was developed and tested on a MacBook (Apple M3, 16\,GB) under Python 3.13 with \texttt{numpy}, \texttt{scipy}, \texttt{scikit-learn}, \texttt{statsmodels}, \texttt{matplotlib}, \texttt{h5py}, and \texttt{bdpy}; per-subject CAV/NVS extraction is parallel-friendly and was run on the same machine. Re-rendering main and appendix figures from the cached JSON outputs takes a few minutes; re-fitting the full per-subject CAV/NVS grid from raw GOD data with the included caches takes on the order of half a day end to end.

\paragraph{Random seeds.} All stochastic components --- the permutation null underlying NVS (default $1{,}000$ random permutations of stimulus pairings), the bootstrap CIs in App.~\ref{app:fw-viability}, the synthetic-PoC random factor draw, and the random-orthogonal projection $A_B$ (\S\ref{sec:poc}) --- are seeded deterministically (\texttt{numpy.random.default\_rng(seed)} with seed $42$ for the main reported results) so that the released code reproduces the reported figures and tables up to numerical / BLAS-level variation.

\end{document}